\documentclass[%
 reprint,
nofootinbib,
 amsmath,amssymb,
 aps,
 prd,
]{revtex4-1}

\usepackage{graphicx}
\usepackage{dcolumn}
\usepackage{bm}
\usepackage[linktocpage,pdfusetitle,colorlinks]{hyperref}

\DeclareFontFamily{OT1}{pzc}{}
\DeclareFontShape{OT1}{pzc}{m}{it}{<-> s * [1.10] pzcmi7t}{}
\DeclareMathAlphabet{\mathpzc}{OT1}{pzc}{m}{it}
\usepackage[normalem]{ulem}

\usepackage{amsmath}
\usepackage{siunitx}
\DeclareSIUnit[]\sunmass{\text{\ensuremath{M_{\odot}}}}

\usepackage{mathtools}

\DeclarePairedDelimiter\abs{\lvert}{\rvert}%
\DeclarePairedDelimiter\norm{\lVert}{\rVert}%

\usepackage{pgfplots}

\pgfplotsset{compat=newest}
\usepgfplotslibrary{groupplots}

\usepackage{enumitem}
\usepackage{bbm}

\newlength\figureheight
\newlength\figurewidth
\makeatletter
\let\oldabs\abs
\def\abs{\@ifstar{\oldabs}{\oldabs*}}
\let\oldnorm\norm
\def\norm{\@ifstar{\oldnorm}{\oldnorm*}}
\makeatother

\renewcommand{\d}[1]{\ensuremath{\operatorname{d}\!{#1}}}
\renewcommand{\vec}[1]{\ensuremath{\mathbf{#1}}}

\begin{document}

\title{Initial data for general relativistic simulations of multiple electrically charged
  black holes with linear and angular momenta}

\author{Gabriele Bozzola}
\email{gabrielebozzola@email.arizona.edu}
\affiliation{Department of Astronomy, University of Arizona, Tucson, AZ, USA}
\author{Vasileios Paschalidis}
\email{vpaschal@email.arizona.edu}
\affiliation{Departments of Astronomy and Physics, University of Arizona, Tucson, AZ, USA}

\date{\today}

\begin{abstract}
  A general relativistic, stationary, and axisymmetric black hole in a
  four-dimensional asymptotically-flat spacetime is fully determined by its
  mass, angular momentum and electric charge. The expectation that
  astrophysically relevant black holes do not posses charge has resulted in a
  limited number of investigations of moving and charged black holes in the
  dynamical, strong-field gravitational (and electromagnetic) regime, in which
  numerical studies are necessary. Apart from having a theoretical interest, the
  advent of multimessenger astronomy with gravitational waves offers new ways to
  think about charged black holes. In this work, we initiate an exploration of
  charged binary black holes by generating valid initial data for general
  relativistic simulations of black hole systems that have generic electric
  charge, linear and angular momenta. We develop our initial data formalism
  within the framework of the conformal transverse-traceless (Bowen-York)
  technique using the puncture approach, and apply the theory of isolated
  horizons to attribute physical parameters (mass, charge, and angular momentum)
  to each hole. We implemented our formalism in the case of a binary system by
  modifying the publicly available \texttt{TwoPunctures} and
  \texttt{QuasiLocalMeasures} codes. We demonstrate that our code can recover
  existing solutions and that it has excellent self-convergence properties for a
  generic configuration of two black holes.
\end{abstract}

\maketitle

\section{Introduction}
\label{sec:introduction}

The successful detection of gravitational waves from the inspiral and merger of
binary black holes by the LIGO-Virgo interferometers~\cite{Abbott2016b,
  Abbott2016c, Abbott2017d, Abbott2017e, LIGOVirgo2018} was made possible not
only by technological advancements in instrumentation but also by substantial
improvements in theoretical modeling that furnished the gravitational wave
templates necessary for performing matched filtering \cite{Flanagan1998a,
  Flanagan1998b, Aylott2009, Ajith2012,Hinder:2013oqa}. To generate a bank of
complete template signals, the equations of general relativity have to be solved
during the late compact binary inspiral, merger and post-merger phases, because
these events involve extreme gravitational fields, whose description with
post-Newtonian methods is not accurate. Obtaining an analytic solution to
describe these systems during these dynamic stages is not possible. Therefore,
numerical integration of the full Einstein equations provides the only viable
avenue for understanding such relativistic astrophysical systems from first
principles and for helping to build gravitational wave templates during the
most dynamical phases of their evolution.

Assuming that general relativity is the correct theory of gravity, the problem
of two black holes is solved by integrating Einstein's equations in vacuum.
Despite the simpler description of black hole spacetimes compared to spacetimes
with matter, it took decades for the field of numerical relativity to mature
enough to be able to stably evolve two black holes until
merger~\cite{Pretorius:2005gq, Campanelli:2005dd, Baker:2005vv}. Some of the
issues that hindered the development were due to the highly non-linear character
of the Einstein equations, the coordinate freedom of general relativity, and the
intrinsically singular nature of black holes. However, since the 2005
breakthrough, numerical relativity has advanced considerably with
state-of-the-art codes that can simulate the inspiral and merger of uncharged
binary black holes, and extract gravitational waves with high precision (see,
e.g.,~\cite{Gopakumar:2007vh,Lovelace2012,Chu:2015kft,Lovelace:2016uwp, Jani2016,
  Healy2017, Healy2019} and references therein). Numerical relativity furnishes
invaluable information for gravitational-wave detection and analysis, which
includes the development of templates (see,
e.g.,~\cite{Baumgarte2008,Scheel2009,Ajith2012,Hinder:2013oqa}), and the accurate
parameter estimation of already detected events \cite{Lange:2017wki}.

Apart from binary black holes, binary neutron stars and binary black
hole--neutron stars are also most the promising gravitational wave sources for
currently operating interferometers~\cite{BSBook}. In fact, among the eleven
confirmed detections of gravitational waves so far \cite{LIGOVirgo2018}, event
GW170817 is attributed to the inspiral and merger of a binary neutron star
\cite{Abbott2017a} (although a binary black hole--neutron star cannot be ruled
out~\cite{Yang:2017gfb,Hinderer:2018pei,Foucart:2018rjc,Coughlin:2019kqf} as a
possibility). A complete simulation of compact binaries with matter requires the
evolution of the spacetime coupled to matter, radiation/neutrinos, and
electromagnetic fields in conjunction with detailed microphysics. A full
solution including radiation/neutrinos without approximation is impossible at
this time, and even with approximation, evolution of perfect fluids with
existing numerical schemes involves density floors and other ad hoc
prescriptions that are necessary to stabilize the calculations (see,
e.g.,~\cite{Font2008,Etienne:2011ea}), but are designed such that their impact on
the global solution is minimal. However, this means that in a sense, simulations
involving perfect fluids are not as ``clean'' as the ones in vacuum, which do
not require ad hoc prescriptions. Nevertheless, many important results have been
obtained through binary neutron star and binary black hole--neutron star
simulations in full general relativity,
see~\cite{Shibata2011,Faber2012,Lehner:2014asa,Paschalidis2017,PaschalidisStergioulas2017,Baiotti2017,Duez:2018jaf}
for reviews (see also~\cite{Cardoso:2012qm} for other applications of
numerical relativity).

Interesting spacetimes that are as ``clean'' as vacuum spacetimes, but
have received little attention in numerical relativity, are those
described by Einstein-Maxwell's theory. This theory involves only
gravitational and electromagnetic fields, and the corresponding
spacetimes are referred to as \emph{electrovacuums} or
\emph{electrovacs}. However, force-free electrodynamics has received
some attention \cite{Palenzuela2010, Lehner2011, Moesta2011,
  Alic2012,Paschalidis:2013gma, Paschalidis2013, Ponce2014, East2018},
but those simulations are not ``clean'', in the sense that when the
force-free conditions are violated during the evolution (typically in
current sheets), one must interfere and enforce them to continue the
calculations. On the other hand, electrovacuum spacetimes can be
solved without physical approximations or ad hoc prescriptions, as the
only assumption here is that electromagnetism and gravitation are
described by the source-free Einstein-Maxwell equations. This
simplification is the reason why these spacetimes have attracted
numerous theoretical and analytic investigations for a long time,
including the celebrated Kaluza-Klein theory \cite{Kaluza1921,
  Klein1926} unifying gravity and electromagnetism.

Examples of interesting electrovacuum spacetimes are those with electrically
charged black holes.\footnote{It is also possible to include magnetic charges.
  This will not be done in the study presented in this paper, so we always take
  the term \emph{charge} to mean \emph{electric charge}. We note the extension
  of the work to include magnetic charges would be straightforward.} The case of
a single charged non-rotating black hole is analytically solved by the
Reissner-Nordstr{\"o}m metric \cite{Reissner1916, Nordstrom1918}. This solution has
been extended to non-vanishing angular momentum in the Kerr-Newman spacetime
\cite{Newman1965}, which generalizes the uncharged rotating black hole solution
found by Kerr \cite{Kerr1963}. Another interesting class of solutions with
multiple black holes is the static Majumdar-Papapetrou solution
\cite{Majumdar1947, Papapetrou1945} that describes non-spinning black holes
whose electric repulsion and gravitational attraction balance, producing a zero
net force condition, and thus equilibrium. The hypothesis of staticity was
relaxed to simple stationarity by \cite{Perjes1971, Israel1972, Hartle1972}.
This list summarizes the known analytical solutions of the source-free
Einstein-Maxwell equations in four-dimensional asymptotically-flat spacetimes.

A reason why the source-free Einstein-Maxwell theory has been primarily confined
to the realm of theoretical explorations is the fact that astrophysically
relevant black holes are not believed to be electrically charged, as the charge
would be neutralized by the surrounding plasma \cite{Wald1984} or as result of a
pair-production through a Schwinger-like process \cite{Gibbons1975}.
Nonetheless, there are some viable mechanisms that result in a black hole with non-zero
charge. One example is the model proposed by \cite{Wald1974}, where the charge
is retained due to the presence of an external magnetic field. This is known as
the ``Wald mechanism''. It was shown in \cite{Wald1974} that if an
asymptotically uniform magnetic field $B_0$ can be sustained, a black hole with
mass $M$ spinning with angular momentum $J$ would acquire an electric charge
$Q = 2 B_0 J$ (measured in geometrized units\footnote{The conversion factor
  between our units and the International System of Units (SI) is
  $c^2 G^{-\frac{1}{2}} (4 \pi \varepsilon_0)^{\frac{1}{2}} =
  \SI{1.16e20}{\coulomb\per\kilo\m}$, so
  $\SI{1}{\sunmass} = \SI{1.71e20}{\coulomb}$, with $c$ speed of light in
  vacuum, $G$ gravitational constant, $\varepsilon_{0}$ vacuum permittivity and
  $M_\odot$ solar mass.}), which we can rewrite as $Q/M = 2 B_0 \chi M$ with
$\chi=J/M^2$ the black hole dimensionless spin parameter. Since for black holes
$\chi^2 \leq 1$, there exists a maximum possible charge-to-mass ratio in the Wald
mechanism: $(Q/M)\leq (Q/M)_{\rm max}\equiv 2B_0 M$~\cite{Wald1974}. In the case of a
solar mass black hole in the galactic magnetic field
\cite{Eckart2012,Eatough2013} the ratio has to be $Q \slash M \le \num{e-24}$. The
charge-to-mass ratio quantifies the deformation of the spacetime due to
electromagnetism, so if it is very small it means that the spacetime is
well-described by a vacuum (uncharged) black hole. Black holes with mass
$M \gtrsim \SI{e9}{\sunmass}$ immersed in a magnetic field of order
$\num{e11}~\mathrm{G}$ would be needed to reach values of $Q \slash M$ large enough
to be relevant for the spacetime structure. Fields of such strength are expected
to be found only in neutron stars. Based on the Wald mechanism, it has been
recently proposed that a binary black hole -- neutron star could provide a
suitable environment to charge the black hole itself \cite{Levin2018}. A second
case in which charged black holes might occur in the Universe is immediately
after the collapse of a compact star when the resulting hole might briefly
retain some charge \cite{Ray2003}. A similar scenario is the collapse of
magnetized stars \cite{Nathanail2017}, which was also considered as a candidate
for fast-radio bursts \cite{Liu2016}. Finally, charged black holes can emerge
in more exotic theories associated with ``hidden'' gauge fields and elementary particles
whose charge is a fraction of the electron charge~\cite{Cardoso2016b}.

In spite of the apparently compelling reasons to believe that
astrophysical black holes have practically zero net charge compared to
their mass, it is still worth studying the source-free
Einstein-Maxwell system to advance our comprehension of strong-field
gravitation and electromagnetism in this largely unexplored
territory. The interplay between electromagnetism and gravity in a
highly dynamical spacetime, which can be probed only with numerical
investigations, can offer a unique laboratory for both theoretical and
more exotic astrophysical studies. For example, the inclusion of
charge in highly relativistic collisions of black holes (see,
e.g.,~\cite{Sperhake2008, Sperhake2009, Berti2010, Sperhake2013} for
such studies with zero charge) would advance our understanding in a
new direction never explored before. Another interesting application
of dynamical electrovacuums is related to cosmic censorship. In a
recent series of papers, it was argued that strong cosmic censorship
can be violated by electrovacuums with a positive cosmological
constant \cite{Mo2018, Dias2018, Dias2018b}. In contrast, the case
without cosmological constant is not settled yet \cite{Cardoso2018}.

The coalescence and merger of charged black holes may present new interesting
phenomenology. For instance, \cite{Zhang2016} proposed that the faint potential
electromagnetic counterpart to GW150914 \cite{Abbott2016, Connaughton2016} might
have been the result of merger of charged black holes. Another hypothesized
mechanism along similar lines invokes magnetic reconnection
\cite{Fraschetti2018}. Subsequently, \cite{Liebling2016} tested the idea of
\cite{Zhang2016} with relativistic simulations. However, the set-up considered
by the authors had some limitations: only equal-mass, equal-charge, non-rotating
black holes were studied, and the initial data did not satisfy the constraints
of the field equations. A more systematic study of this part of the parameter
space of charged black holes requires that one starts with constraint-satisfying
initial data for black hole configurations with arbitrary charge, mass ratio,
linear and angular momenta.

The most common avenue for generating constraint-satisfying initial data is
provided by the $3+1$ decomposition of spacetime
\cite{Arnowitt:1962hi,York1971}. In this approach, one casts the
Einstein-Maxwell equations to an initial value problem in which the
four-dimensional spacetime is foliated by successive timeslices obtained via the
dynamical evolution of the system.\footnote{It is worth mentioning that another
  common approach to building spacetimes in the computer is the
  \emph{generalized harmonic formalism}
  \cite{Pretorius:2004jg,Pretorius:2005gq}.} When performing this
decomposition, both Maxwell's and Einstein's equations are split in two sets:
the evolution and the constraint equations. The former move the system forward
in time, whereas the latter must be satisfied at all times and must be used to
generate the initial data for the evolution. In this paper, we primarily focus on
the constraint equations.

Einstein-Maxwell's theory was first cast in a $3+1$ decomposition by
\cite{Thorne1982} and more than 25 years later, \cite{Alcubierre2009} proved
that the evolution equations are symmetric hyperbolic, and hence admit a
well-posed initial value problem. Moreover, the authors extended the work of
\cite{Bowen1985} to generate initial data for electrically charged black holes
at a moment of time-symmetry -- the spacetime is invariant with respect to
time reversal. Recent applications of this formalism are the head-on collisions
by \cite{Zilhao2012, Zilhao2013} (the interested reader can find several cogent
additional reasons motivating the numerical study of charged black holes in
these references). In these works, the authors evolved initial data generated
with the same formalism described by \cite{Alcubierre2009} and were mostly
interested in comparing the electromagnetic and gravitational emissions.
Finally, the same group also investigated numerically the non-linear stability of
a Kerr-Newman black hole \cite{Zilhao2014}.

In this paper, we extend the work of \cite{Alcubierre2009} and \cite{Zilhao2014}
to generate initial data for charged, rotating, and moving black holes in a
self-consistent way.\footnote{We note that the formalism outlined in this paper
  applies not only to electromagnetism but to any U(1) charge (such as the one
  described in \cite{Cardoso2016b}).} We adopt the \emph{conformal
  transverse-traceless} formalism \cite{Bowen:1980yu} treating the black holes
as punctures to solve for the metric, and take advantage of the
Reissner-Nordstr{\"o}m solution in isotropic coordinates to solve for the
electromagnetic fields. This strategy involves two major challenges. The first
is that non-linear partial differential equations have to be solved. This can be
done only numerically for generic binary black hole configurations. To address
this issue we modify the \texttt{TwoPuncture} code \cite{Ansorg2004} to solve
the resulting elliptic differential equations. The single-domain pseudo-spectral
character of the code results in an accurate solution and it is quickly
convergent. The second challenge is that the physical interpretation of the
results is not transparent. The parameters given as input for the algorithm (the
\emph{bare parameters}, such as mass and charge) in general are not actual
\emph{physical} quantities of the resulting black holes. Hence, we apply the
theory of isolated horizons \cite{Ashtekar2000}, which provides a quasi-local
machinery for linking the bare black hole parameters with the physical ones and
is suitable for simulations. We implement this numerically by modifying the
\texttt{QuasiLocalMeasures} code \cite{Dreyer2003}.

We structure the paper as follows. In Section~\ref{sec:formalism} we review the
mathematical tools necessary for generating initial data for charged black
holes. In particular, we present the $3+1$ decomposition of Einstein-Maxwell's
theory and review the Reissner-Nordstr{\"o}m solution in isotropic coordinates and
the formalism of isolated horizons. In Section~\ref{sec:solv-constr-equat} we
solve the constraints with the conformal transverse-traceless technique. Our
numerical implementation and tests are detailed in Section~\ref{sec:numer-impl}.
Finally, Section~\ref{sec:concl-future-work} summarizes our findings and
describes possible future research directions.

In Appendix~\ref{sec:algor-import-equat}, we prepared a summary of the
important equations and steps needed to generate initial data for
generic systems of charged black holes. The Appendix provides a
distilled overview of the analytic content of this paper. For the
reader who is interested only in the gist of the algorithm/equations
and the results of our work, we suggest they skip to
Appendix~\ref{sec:algor-import-equat}, and then read
Sections~\ref{sec:code-validation} and~\ref{sec:convergence}, where we
present our results.

\subsubsection*{Notation and conventions}
\label{sec:notation-conventions}

We assume that gravity and electromagnetism are described by Einstein-Maxwell's
theory \cite{Wald1984} and we follow the same notation as in \cite{MTW1973}. In
particular, we use Einstein's summation convention and the signature of the
metric is $(-, +, +, +)$. We use geometrized units with $G=c=1$, where $c$ is
the speed of light in vacuum and $G$ is the gravitational constant. The unit of
charge is defined so that the proportionality constant in Coulomb's law is 1
(for more details, see \cite{Jackson1975}). Indices $a$, $b$, $c$, and $d$ run
in the set $\{0,1,2,3\}$, whereas the other Latin letters, such as $i$, $j$ or
$k$, run in the set $\{1,2,3\}$ and are referred to as spatial
components. Parentheses and square brackets in the indices mean symmetrization and
anti-symmetrization, respectively. We also use the abstract index notation
\cite{Wald1984}. We reserve the symbol $\nabla$ for the four-dimensional covariant
derivative associated with the spacetime metric $g_{ab}$, and $D$ for the
three-dimensional covariant derivative, compatible with the spatial metric
$\gamma_{ij}$. We denote the determinant of these metrics as $g = \det{g_{ab}}$ and
$\gamma = \det{\gamma_{ij}}$. We prepend the symbol ``${(4)}$'' to all the
four-dimensional tensors, with exception of the metric $g_{ab}$. For the
completely antisymmetric Levi-Civita tensor we use the convention that
$\epsilon_{1230} = \sqrt{-g}$, and $\epsilon_{123} = \sqrt{\gamma}$, and denote the Levi-Civita
symbol with $\bar{\epsilon}_{ijk}$ or $\bar{\epsilon}^{ijk}$.

\section{Formalism}
\label{sec:formalism}

In this Section we describe the theoretical tools that we use later to
generate initial data for arbitrary configurations of charged black
holes.  Specifically, in Section~\ref{sec:einstein-maxwell-3+1} we
survey the $3+1$ decomposition of Einstein-Maxwell's equations,
focusing on the constraint
equations. Section~\ref{sec:Reissner-Nordstrom} reviews the
Reissner-Nordstr{\"o}m solution for a single charged stationary black hole
in isotropic coordinates.  Section~\ref{sec:dynamical-horizons}
summarizes the theory of isolated horizons, which we employ to assign
the black hole physical properties: mass, charge and angular momentum.

\subsection{$3+1$ decomposition of Einstein-Maxwell}
\label{sec:einstein-maxwell-3+1}

In this paper we study systems described by the source-free
Einstein-Maxwell equations \cite{Wald1984}
\begin{subequations}
    \label{eq:einstein-equations}
  \begin{align}
    {}^{(4)}\mathcal{R}_{ab} - \frac{1}{2}g_{ab} {}^{(4)}\mathcal{R}  &=
     8 \pi {}^{(4)}T^{\mathrm{EM}}_{ab}   \,, \\
    \nabla_a {}^{(4)}F^{ab} &= 0\,, \\
    \nabla_a {}^{(4)}{}^{\star}F^{ab} &= 0 \,,
  \end{align}
\end{subequations}
where ${}^{(4)}\mathcal{R}_{ab}$ is the Ricci tensor associated with the metric
$g_{ab}$, ${}^{(4)}\mathcal{R}={}^{(4)}\mathcal{R}^{a}_{\; a}$,
${}^{(4)}F_{ab} = 2\, {}^{(4)}A_{[a,b]}$ is the Maxwell field-strength tensor, with
${}^{(4)}A_a$ the four-potential, and ${}^{(4)}{}^{\star}F_{ab}$ is its Hodge dual,
defined by
\begin{equation}
\label{eq:fstar}
{}^{(4)}{}^{\star}F^{ab} = \frac{1}{2} \epsilon^{abcd} \,{}^{(4)}F_{cd}\,.
\end{equation}
The electromagnetic stress-energy tensor is
\begin{equation}
  \label{eq:electromagnetic-stress-energy-tensor}
  4 \pi {}^{(4)}T_{ab}^{\mathrm{EM}} =  {}^{(4)}F_{ac} {}^{(4)}F_{bd} g^{cd} -
  \frac{1}{4}g_{ab} {}^{(4)}F_{cd} {}^{(4)}F^{cd}\,.
\end{equation}
Solving the coupled Einstein-Maxwell equations in four dimensions is a
challenging task. In particular, the form of
Equations~\eqref{eq:einstein-equations} is not suitable for a numerical
solution. Therefore, we adopt the standard $3+1$ decomposition to express the
equations as a Cauchy problem, and cast them in a form amenable for numerical
integration~\cite{Baumgarte:2010nu}.

Assuming that the spacetime is described by a globally hyperbolic
Lorentzian manifold $\mathcal{M}$ with metric tensor $ g_{ab}$,
$\mathcal{M}$ can be foliated by a family of spacelike
non-intersecting hypersurfaces $\Sigma_t$, taken as level surfaces of
a time function $t$. Let $n^a$ be the future-directed, timelike unit
vector normal to $\Sigma_t$. The projection operator along this vector
is $n^a n_b$, whereas the one onto $\Sigma_t$ is
\begin{equation}
  \label{eq:projector}
  \gamma^a_{\; b} = \delta^{a}_{\; b} + n^a n_b\,.
\end{equation}
The induced metric on $\Sigma_t$, is derived by applying twice the
projection operator on $g_{ab}$, which yields
\begin{equation}
  \label{eq:three-metric}
  \gamma_{ab} = g_{ab} + n_a n_b\,.
\end{equation}
The induced metric is purely spatial ($\gamma_{ab}n^b=0$), it encodes the intrinsic
curvature of the hypersurfaces $\Sigma_t$ and can be used to define a spatial
covariant derivative $D_i$ on $\Sigma_t$.

Instead of working with the normal vector $n^a$, it is
convenient to use the normalized time vector
\begin{equation}
  \label{eq:time-vector}
  t^a = \alpha n^a + \beta^a\,,
\end{equation}
where $\alpha$ and $\beta^a$ are the \emph{lapse function} and \emph{shift vector}.
With these quantities, the spacetime metric assumes the
Arnowitt-Deser-Misner (ADM) form~\cite{Arnowitt:1962hi,Arnowitt2008}
\begin{equation}
  \label{eq:metric-3-1}
  \d s^2 = - \alpha^2 \d t^2 + \gamma_{ij} (\d x^i + \beta^i \d t)(\d x^j + \beta^j \d t)\,.
\end{equation}

The spatial metric is not sufficient to fully describe the curvature
properties of the four-dimensional spacetime. The extrinsic curvature
$K_{ab}$ supplies the missing information by expressing how $\Sigma_t$ is
embedded in $\mathcal{M}$, and is  defined as
\begin{equation}
  \label{eq:extrinsic-curvature}
  K_{ab} = - \gamma_a^{\; c} \gamma_b^{\; d} \nabla_c n_d\,.
\end{equation}

Just like the induced metric (which we will also refer to as the three-metric
throughout), the extrinsic curvature is purely spatial. The Riemann tensor can
be expressed in terms of $\gamma_{ij}$ and $K_{ij}$, and therefore Einstein's
equations can be rewritten in terms of $3+1$ quantities. The resulting 3+1 ADM
(\emph{{\`a} la} York) formalism \cite{Arnowitt:1962hi, York1971} of general
relativity consists of four constraints and twelve evolution equations. The
constraints are the direct consequence of the integrability conditions that
$\gamma_{ij}$ and $K_{ij}$ have to satisfy to have $\Sigma_t$ properly embedded in
$\mathcal{M}$. On the other hand, the evolution equations provide a prescription
to move from one timeslice to the next provided a gauge choice is made. The
evolution equations preserve the constraints: if the constraints are initially
satisfied, they will always be satisfied. However, when they are not satisfied,
the simulated system is not a solution of the Einstein equations. The same split
into evolution equations and constraint equations holds for Maxwell's theory,
too. In complete analogy to Einstein's theory, Maxwell's evolution equations
preserve the Maxwell constraints, if the constraints are initially satisfied.
For this reason, it is important to start with valid, constraint-satisfying
initial data. In this work, we focus only on the constraint equations, precisely
because our goal is the generation of valid initial data for general
relativistic simulations in Einstein-Maxwell theory.

Let ${}^{(4)}T_{ab}^{\text{EM}}$ be the stress-energy tensor, and define
\begin{subequations}
  \label{eq:constraint-sources}
  \begin{eqnarray}
  \mathcal{E} &= n_a n_b {}^{(4)}T^{ab}_{\text{EM}}\,,  \\
  S^i &= - \gamma^{ij} n^a {}^{(4)}T_{aj}^{\text{EM}}\,.
  \end{eqnarray}
\end{subequations}
The Einstein constraints then become~\cite{Baumgarte:2010nu}
\begin{subequations}
    \label{eq:constraints-ADM}
  \begin{eqnarray}
    \label{eq:hamiltonian-ADM}
    \mathcal{R} + K^2 - K_{ij} K^{ij} &= 16 \pi \mathcal{E} \,, \\
    \label{eq:momentum-ADM}
    D_j(K^{ij} - \gamma^{ij} K) &= 8 \pi S^i \,,
  \end{eqnarray}
\end{subequations}
with $\mathcal{R}$ being three-dimensional Ricci scalar associated with
$\gamma_{ij}$, and $K$ the trace of the extrinsic
curvature. Equation~\eqref{eq:hamiltonian-ADM} is known as the
\emph{Hamiltonian} constraint, Equations~\eqref{eq:momentum-ADM} as the
\emph{momentum} constraints.

Equations~\eqref{eq:constraints-ADM} are not the only constraints in
Einstein-Maxwell's theory. As for Einstein's equations, a $3+1$ split of
Maxwell's equations must be performed.\footnote{A more detailed derivation of
  the three-dimensional Maxwell equations from the four-dimensional ones can
  be found in the Appendix of \cite{Alcubierre2009} (see also
  \cite{Baumgarte2003b}).} First, we
introduce the electric and magnetic fields as measured by normal observers
with four-velocity $n^a$,
\begin{subequations}
  \label{eq:em-norma}
  \begin{align}
    E^a & = {}^{(4)}F^{ab} n_b\,, \\
    B^a  & = {}^{(4)}{}^{\star}F^{ab} n_b = \frac{1}{2} \epsilon^{abcd} n_b {}^{(4)}F_{cd}\,,
  \end{align}
\end{subequations}
which are both purely spatial ($n_a E^a=n_a B^a = 0$). The electromagnetic
tensor becomes
\begin{equation}
  \label{eq:three-faraday-tensor}
  {}^{(4)}F_{ab} = n_a E_b - n_b E_a +  \epsilon_{abcd} B^c n^d\,,
\end{equation}
and its dual is
\begin{equation}
  \label{eq:three-hodge-faraday-tensor}
  {}^{(4)}{}^{\star}F_{ab} = n_a B_b - n_b B_a - \epsilon_{abcd} E^c n^d\,.
\end{equation}
With these decompositions, Maxwell's equations can be expressed in
terms of $3+1$ quantities. As in the case of the Einstein equations,
the $3+1$ split leads to evolution and constraint equations. In
particular, the electromagnetic constraints are
\begin{subequations}
  \label{eq:maxwell-contraints}
  \begin{align}
    D_a E^a &= 0 \,,\\
    D_a B^a &= 0\,.
  \end{align}
\end{subequations}
The electromagnetic sector couples with the spacetime through the
stress-energy tensor $T^{ab}_{\mathrm{EM}}$ which is re-written in terms
of the 3+1 variables as
\begin{multline}
  \label{eq:stress-energy-3d-em}
  4 \pi T^{ab}_{\mathrm{EM}} = \frac{1}{2} (n^a n^b + \gamma^{ab})(E_c E^c + B_c B^c)\
  \\ + 2 n^{(a} \epsilon^{b)cd} E_c B_d - (E^a E^b + B^a B^b) \,,
\end{multline}
where $\epsilon^{bcd} = n_a \epsilon^{abcd}$. Plugging Equation~\eqref{eq:stress-energy-3d-em}
into the source terms of Equations~\eqref{eq:constraint-sources}, we find
\begin{subequations}
  \label{eq:em-sources}
  \begin{align}
    \label{eq:em-sources-rho}
    4\pi \mathcal{E} &= \frac{1}{2} (E_i E^i + B_i B^i)\,, \\
    \label{eq:em-sources-S}
    4\pi S^i &= \epsilon^{ijk} E_j B_k\,,
  \end{align}
\end{subequations}
which are the familiar electromagnetic energy density and Poynting vector.

\subsection{The Reissner-Nordstr{\"o}m spacetime}
\label{sec:Reissner-Nordstrom}

The Reissner-Nordstr{\"o}m spacetime \cite{Reissner1916, Nordstrom1918}
describes an isolated non-rotating black hole with electric charge $q$
and mass $m$ \cite{Wald1984}. This solution will be the base of our
generalization to charged black hole systems. In Boyer-Lindquist
coordinates ($t, r, \theta, \phi$), the Reissner-Nordstr{\"o}m metric is given by
\begin{multline}
  \label{eq:rn-usual-metric}
  \d s^2 = - \left(1 - \frac{2m}{r} + \frac{q}{r^2}\right) \d t^2
  + \\
  \left(1 - \frac{2m}{r} + \frac{q}{r^2}\right)^{-1} \d r^2 + r^2 (\d \theta^2 +
  \sin^2 \theta \d \phi^2)\,,
\end{multline}
and the electromagnetic potential of the solution is
\begin{equation}
  \label{eq:rn-A}
  {}^{(4)}A = -\frac{q}{r} \d t\,.
\end{equation}
In the following Sections we will adopt the puncture approach, so we transform
the Boyer-Lindquist coordinates to isotropic ones. In order to do so, we define
a new radial coordinate $R$ satisfying
\begin{equation}
  \label{eq:isotropic-coordinates}
  r = R \left(1 + \frac{m}{R} + \frac{R_H^2}{R^2}\right)\,,
\end{equation}
with $R_H = \frac{1}{2} \sqrt{m^2 - q^2}$ the radius of the black hole
horizon in isotropic coordinates. The metric then assumes the following form
\begin{equation}
  \label{eq:RN-isotropic}
  \d s^2 = -\Psi^{-4} \d t^2 + \Psi^{4} \delta_{lk} \d x^l \d x^k\,,
\end{equation}
with $\delta_{lk}$ the flat Euclidean metric, and $\Psi$ the conformal factor
defined as
\begin{equation}
  \label{eq:RN-Psi}
  \Psi = \sqrt{1 + \frac{m}{R} + \frac{R_H^2}{R^2}} =  \sqrt{\left(1 + \frac{m}{2R}\right)^2 - \left(\frac{q}{2R}\right)^2}\,.
\end{equation}
As is clear from Equation~\eqref{eq:RN-isotropic}, the spatial metric
is manifestly conformally flat in isotropic coordinates. Moreover,
there is no magnetic field and the electric field has only an $R$
component
\begin{equation}
  \label{eq:ER-KN}
  E^R = \Psi^{-6} \frac{q}{R^2}\,.
\end{equation}
As a result, the Poynting vector defined in
Equation~\eqref{eq:em-sources-S} is identically zero
everywhere.

\subsection{Isolated horizons}
\label{sec:dynamical-horizons}

Once the constraint equations are solved, it is important to interpret the
physical configuration to which the initial data correspond. This can be achieved by
locating the black hole apparent horizons and applying the theory of
\emph{isolated horizons}~\cite{Ashtekar2000} (see
\cite{Ashtekar2004} for a review). Isolated horizons provide a quasi-local notion of the
black hole physical properties. In this Section we review basic identities of
the formalism, including, in particular, the electric charge of the horizon, and
the electromagnetic field contribution to angular momentum, elements that have
not received much attention in numerical relativity
applications~\cite{Dreyer2003, Schnetter2006}.

Isolated horizons have several desirable features. For instance, they always lie
inside the event horizon, to which they reduce for stationary spacetimes, and
they imply the existence of a future singularity \cite{Penrose1965,
  Hawking1970}. Most relevant for our purpose, they provide well-defined notions
of mass, charge and angular momentum. For spacetimes with suitable symmetries,
these quasi-local physical quantities coincide with the global ones defined from
conservation laws (for example via ADM integrals), as we verify this explicitly
for the Reissner-Nordstr{\"o}m case in Appendix~\ref{sec:reissner-nordstrom-1}.
However, in general, the quasi-local definitions and those at infinity differ
\cite{Ashtekar2000}. Furthermore, the formalism does not provide a quasi-local
definition of linear momentum due to the lack of a meaningful notion of
space-translational symmetry in curved spacetime
\cite{Ashtekar2004,Krishnan2002}.

Here, we follow closely \cite{Dreyer2003} in using isolated horizons to assign black
hole physical parameters. Given a spatial section $\mathcal{S}$ of an isolated
horizon, the variables we are interested in are defined as follows. First, the
areal radius is given by
\begin{equation}
  \label{eq:isolated-radius}
  R_{\mathcal{S}} = \left( \frac{1}{4\pi} \int_{\mathcal{S}} \epsilon \right)^{\frac{1}{2}}\,,
\end{equation}
where $\epsilon$ is the area two-form on the 2-surface, given by
$\epsilon = \frac{1}{2} \sqrt{\mathbbm{q}} \bar{\epsilon}_{ab} \d x^a \wedge \d x^b$, where
$\mathbbm{q}_{ab}$ is the induced metric on the horizon,
$\mathbbm{q} = \det \mathbbm{q}_{ab}$, and $\bar{\epsilon}_{ab}$ is the two-dimensional
antisymmetric symbol. $\int_{\mathcal{S}} \epsilon$ is the surface area of the horizon.

Next, the definition of the angular momentum is based on an
approximate rotational killing vector field $\varphi^a$ on the
2-surface
\cite{Ashtekar2000}
\begin{equation}
  \label{eq:isolated-spin}
  J_{\mathcal{S}} = - \frac{1}{8\pi} \int_{\mathcal{S}} (\varphi \cdot \omega)\epsilon + 2(\varphi \cdot {}^{(4)}A) {}^{(4)}{}^{\star}F \,,
\end{equation}
where $\omega$ is the form that satisfies the condition
$t^a \nabla_a k^b = t^a \omega_a k^b$ for any vector $t^a$ tangent to
$\mathcal{S}$, with $k^b$ being the outgoing future-directed vector normal to
$\mathcal{S}$ . By construction of $k^b$, $\omega$ always exists
\cite{Ashtekar2000}. The two terms in the right-hand-side of
Equation~\eqref{eq:isolated-spin} are the gravitational and electromagnetic
contribution to the horizon angular momentum.

The charge is defined by means of Gauss's law
\begin{equation}
  \label{eq:isolated-charge}
  Q_{\mathcal{S}} = \frac{1}{4\pi} \int_{\mathcal{S}} {}^{(4)}{}^{\star}F\,,
\end{equation}
and finally, the gravitational mass of the isolated horizon is given by
\begin{equation}
  \label{eq:isolated-mass}
  M_{\mathcal{S}} = \frac{1}{2R_{\mathcal{S}}} \left[(R_{\mathcal{S}}^2 +
    Q_{\mathcal{S}}^2)^2
    + 4J_{\mathcal{S}}^2\right]^{\frac{1}{2}}\,.
\end{equation}
For Kerr-Newman black holes, this formula perfectly reduces to the equation that
relates total mass, irreducible mass, charge and angular momentum
\cite{Ashtekar2001}.

The definitions of angular momentum and charge involve
four-dimensional quantities, but during simulations with the $3+1$
formalism, it is more convenient to use 3+1 variables. In
\cite{Dreyer2003}, it was shown that the gravitational contribution to
the horizon angular momentum can be computed using an ADM-like formula
\begin{equation}
  \label{eq:isolated-spin-extrinsic-curvature}
  J_{\mathcal{S}}^{\text{GR}} = - \frac{1}{8\pi} \int_{\mathcal{S}} (\varphi \cdot \omega)\epsilon = \frac{1}{8\pi} \int_{\mathcal{S}} \varphi^a R^b K_{ab} \epsilon\,,
\end{equation}
where $R^a$ is the  spatial unit vector normal to $\mathcal{S}$. The
electromagnetic component of the angular momentum
$J_{\mathcal{S}}^{\text{EM}}=J_{\mathcal{S}}-J_{\mathcal{S}}^{\text{GR}}$
depends on both ${}^{(4)}A$ and $ {}^{(4)}{}^{\star}F$. The first is directly
accessible if instead of the electric and magnetic fields one evolves the vector
potential~\cite{DelZanna:2002rv,Giacomazzo:2010bx,Etienne:2011re,Etienne:2015cea,Fragile2018},
whereas the second has components
\begin{equation}
\label{eq::fHF-3-1}
{}^{(4)}{}^{\star}F_{ab}  = (2n_{[a}B_{b]} - \epsilon_{abc} E^c) \,.
\end{equation}
When integrated over a spatial 2-surface the term $2n_{[a}B_{b]}$ does not
contribute because $n_a=(-\alpha,0,0,0)$. Therefore, the electromagnetic contribution to the
horizon angular momentum becomes
\begin{equation}
  \label{eq:isolated-spin-electric}
  J_{\mathcal{S}}^{\text{EM}} = - \frac{1}{4\pi} \int_{\mathcal{S}} (\varphi \cdot {}^{(4)}A) \frac{1}{2!}\epsilon_{abc} E^c \d x^a \wedge \d x^b \,,
\end{equation}
where $\epsilon_{abc}=n^d\epsilon_{abcd}$.
By use of Equation~\eqref{eq::fHF-3-1}, Equation~\eqref{eq:isolated-charge} for the charge becomes
\begin{equation}  \label{eq:isolated-charge-explicit}
  Q_{\mathcal{S}} = \frac{1}{4\pi} \int_{\mathcal{S}} \frac{1}{2!} \epsilon_{abc} E^c \d x^a \wedge \d x^b \,.
\end{equation}
These definitions provide a complete characterization of black holes
during a general relativistic simulation with the $3+1$
decomposition. An example of how the integrations above are performed
is in Appendix~\ref{sec:charge-isol-horiz}.

\section{Solving the constraint equations}
\label{sec:solv-constr-equat}

To solve the constraint equations we adopt the conformal
transverse-traceless approach, also referred to as Bowen-York
technique \cite{Bowen:1980yu}. The goal of this method is to expose
and specify degrees of freedom containing physical information about
the system by applying conformal transformations on the spatial
quantities, and working directly on the \emph{conformal} variables
instead of the \emph{physical} ones.

The first step in the method is to conformally decompose $\gamma_{ij}$  by introducing the
conformal factor $\psi$ and metric $\bar\gamma_{ij}$
\begin{equation}
  \label{eq:conformal-metric}
  \gamma_{ij} = \psi^4 \bar{\gamma}_{ij}\,.
\end{equation}
In the following we use an overbar to indicate conformal quantities.

A common assumption when generating multiple black hole initial data
is that the spatial metric is conformally flat
\cite{Cook2000,Alcubierre2009,Baumgarte:2010nu,Lousto2012}. In other
words, we fix the conformal three-dimensional metric
$\bar{\gamma}_{ij}$ to be the flat Euclidean metric $\delta_{ij}$ (in
Cartesian coordinates).  This choice greatly simplifies computations
and it is a good approximation for the systems we are interested in
studying, in spite of the fact that conformally flat spatial slices of
the Kerr metric do not exist \cite{Garat2000}. Conformal flatness
limits the maximum equilibrium value that the black hole dimensionless
spin can attain \cite{Lovelace2012, Lousto2012}, but values of order
$0.9$ are completely achievable. Thus, we do not anticipate this
approximation to impose severe constraints on the equilibrium values
of the black hole spin and charge.  Considering what happens in the
uncharged case \cite{Brandt1995, Gleiser1998}, we expect that
conformal flatness will generate initial data with spurious
gravitational (and electromagnetic) radiation in the charged black
hole cases, too. Nonetheless, this is not a major concern since in
dynamical simulations the system is evolved until this ``junk''
radiation propagates away, and the fields relax to their
quasi-equilibrium values.

In addition to the conformal decomposition of the metric, it is also
useful to transform the extrinsic curvature $K_{ij}$ by separating it
into its traceless $A_{ij}$ and trace ($K = K^{i}_{\;i}$) parts
\begin{equation}
  \label{eq:K-CTT}
  K_{ij} = A_{ij} + \frac{1}{3}\gamma_{ij} K\,.
\end{equation}
Following standard practice, we adopt the maximal slicing condition
$K = 0$  \cite{Smarr1978}, and introduce a conformal, traceless extrinsic
curvature $\bar A_{ij}$ as
\begin{equation}
  \label{eq:A-conformal}
  K_{ij} = A_{ij} = \psi^{-2} \bar{A}_{ij}\,.
\end{equation}
Then, $\bar{A}_{ij}$ can be split into a transverse-traceless and a longitudinal
part
\begin{equation}
  \label{eq:A-CTT}
  \bar{A}^{ij} = \bar{A}^{ij}_{\text{TT}} + \bar{A}^{ij}_{\text{L}}\,.
\end{equation}
We set $\bar{A}^{ij}_{\text{TT}}=0$, which corresponds to
suppressing the radiative degrees of freedom, so
\begin{equation}
  \label{eq:A-CTT-2}
  \bar{A}^{ij} = \bar{A}^{ij}_{\text{L}}\,.
\end{equation}
The longitudinal part can always be expressed in terms of  a vector
$V$ as
\begin{equation}
  \label{eq:V}
  \bar{A}^{ij} = \bar{A}^{ij}_{\text{L}} = 2 \delta^{ik}\delta^{jh} V_{(h,k)} - \frac{2}{3} \delta^{ij} \partial_k V^k\,,
\end{equation}
where Cartesian coordinates are adopted. Going back to
Equation~\eqref{eq:A-conformal}, the extrinsic curvature is given by
\begin{equation}
  \label{eq:K-traceless-transverse}
  K_{ij} = \psi^{-2} \left( 2 V_{(i,j)} - \frac{2}{3} \delta_{ij} \partial_k V^k\right)\,.
\end{equation}
We already exploited much of the freedom we had in specifying variables during
the previous steps. Under these assumptions, we just need the vector $V^i$ and
the conformal factor $\psi$ to fully determine $\gamma_{ij}$ and $K_{ij}$, and the
constraint Equations~\eqref{eq:constraints-ADM} take the form
\begin{subequations}
  \begin{align}
    \label{eq:hamiltonian-CTT-coupled}
    \nabla^2 \psi + \frac{1}{8} \psi^{-7} \bar {A}_{ij} \bar {A}^{ij} + 2 \pi \psi^{5} \mathcal{E} = 0\,, \\
    \label{eq:momentum-CTT-coupled}
    (\nabla^2 V)^i + \frac{1}{3} \delta^{ij} \partial_j (\partial_k V^k) - 8 \pi \psi^{10} S^i = 0\,,
  \end{align}
\end{subequations}
where $\nabla^2 = \partial_k \partial^k$.

Next, we turn to the electromagnetic sector of the problem. We rescale
the electromagnetic fields as in \cite{Alcubierre2009}
\begin{equation}
  \label{eq:conformally-flat-em-quantities}
  \begin{split}
    \bar{E}^i &= \psi^6 E^i\,, \qquad \bar{E}_i = \psi^2 E_i\,, \\
    \bar{B}^i &= \psi^6 B^i\,, \qquad \bar{B}_i = \psi^2 B_i\,.
  \end{split}
\end{equation}
The factor $\psi^6$ is chosen in order to have
$D_i E^i = \psi^{-6} \partial_i \bar{E}^i$, where we used the fact that for any vector
$v^i$ it holds true that
$D_i v^i = \gamma^{-1\slash2} \partial_i \left( \sqrt{\gamma} v^i \right)$. The Maxwell
constraints~\eqref{eq:maxwell-contraints} read
\begin{subequations}
  \label{eq:conformally-flat-constraints}
  \begin{align}
    \partial_i \bar{E}^i &= 0\,, \\
    \partial_i \bar{B}^i &= 0\,.
  \end{align}
\end{subequations}
These equations do not depend on the conformal factor $\psi$,  so the
electromagnetic constraints can be solved independently from the spacetime ones.
Moreover, the equations are linear; hence we can superpose solutions.

Having fixed the conformal scalings of the $E^i$ and $B^i$ fields,
the source terms $\mathcal{E}$ and $S^i$ of the Einstein constraints
conformally transform as
\begin{subequations}
\begin{align}
  \mathcal{E} &= \psi^{-8} \bar{\mathcal{E}}\,, \\
  S^i & = \psi^{-10} \bar{S}^i\,,
\end{align}
\end{subequations}
where
\begin{subequations}
  \label{eq:em-sources-conformal}
  \begin{align}
    \label{eq:em-sources-rho-conf}
    4\pi \bar{\mathcal{E}} &= \frac{1}{2} (\bar{E}_i \bar{E}^i + \bar{B}_i \bar{B}^i)\,, \\
    \label{eq:em-sources-S-conf}
    4\pi \bar{S}^i &= \bar{\epsilon}^{ijk} \bar{E}_j \bar{B}_k\,.
  \end{align}
\end{subequations}
With these redefinitions, the Einstein constraints become
\begin{subequations}
  \begin{align}
    \label{eq:hamiltonian-CTT}
    \nabla^2 \psi + \frac{1}{8} \psi^{-7} \bar {A}_{ij} \bar {A}^{ij} + 2 \pi \psi^{-3} \bar{\mathcal{E}} = 0\,, \\
    \label{eq:momentum-CTT}
    (\nabla^2 V)^i + \frac{1}{3} \delta^{ij} \partial_j (\partial_k V^k) - 8 \pi \bar {S}^i = 0\,.
  \end{align}
\end{subequations}
The problem is now greatly simplified because the momentum constraints do not
depend on $\psi$, are linear in $V^i$, and along with the Hamiltonian constraint
have decoupled from the Maxwell constraints.

Next, we exploit the linearity of Equation~\eqref{eq:momentum-CTT} by decomposing
$V^i$ as
\begin{equation}
  \label{eq:V-decomposition}
  V^i = V^i_{0,\text{GR}} + V^i_{\text{EM}}\,,
\end{equation}
where $V^i_{0,\text{GR}}$ solves the homogeneous
Equation~\eqref{eq:momentum-ADM} (when $\bar{S}^i = 0$), and ${V}^i_{\text{EM}}$
the inhomogeneous one.\footnote{The subscript $0$ does not indicate any
  component, but it reminder that the field is a solution of the homogeneous
  equation.} The first term does not contain any reference to the
electromagnetic sector of the problem. Thus, as in \cite{Ansorg2004}, we choose
\begin{equation}
  \label{eq:V-GR}
  V^i_{0,\text{GR}} = \sum_{n = 1}^{N_p} \left( - \frac{7}{4} \frac{P^i_n}{R_n}
    - \frac{1}{4} \delta_{jk} x_n^jP_n^k \frac{x_n^i}{R_n^3}
    + \frac{\bar{\epsilon}^i_{\,\,jk} x_{n}^jS_{n}^k}{R_n^3} \right)\,,
\end{equation}
with $R_n = \abs{\vec{x} - \vec{x}_n}$ the Euclidean coordinate
distance from puncture $n$, where $\vec{x}_n$ is the location of the
$n$-th puncture, and $P_n^i$ and $S^k_n$ are its linear and angular
momenta, respectively. Equation~\eqref{eq:V-GR} solves the homogeneous version of
Equation~\eqref{eq:momentum-CTT}, and it is known that for suitable
single black hole solutions $P_{\text{ADM}}^i = P^i$ and
$J_{\text{ADM}}^i = S^i$, with $P_{\text{ADM}}$ and $J_{\text{ADM}}$
being the ADM linear and angular momenta evaluated at infinity
\cite{Bowen:1980yu, Ansorg2004}, respectively. By use of the
decomposition~\eqref{eq:V-decomposition}, the momentum constraints
further reduce to three decoupled linear equations for
$V^i_{\text{EM}}$, effectively replacing
Equation~\eqref{eq:momentum-CTT} with
\begin{equation}
  \label{eq:momentum-CTT-2}
  \nabla^2 V_{\text{EM}}^i + \frac{1}{3} \delta^{ij} \partial_j (\partial_k V^k_{\text{EM}})
  - 8 \pi \bar {S}^i = 0\,.
\end{equation}
We also manipulate the Hamiltonian constraint~\eqref{eq:hamiltonian-CTT} further by
separating the singular part of the conformal factor from the finite one $u$,
motivating our ansatz based on the conformal factor of the Reissner-Nordstr{\"o}m spacetime in
Equation~\eqref{eq:RN-Psi},
\begin{equation}
  \label{eq:psi-singular}
  \psi = \left[ \left(1 + u + \sum_{n=1}^{N_p} \frac{M_n}{2 R_n} \right)^2
    - \left( \sum_{n=1}^{N_p} \frac{Q_n}{2R_n}  \right)^2
  \right]^{\frac{1}{2}}\,.
\end{equation}
We introduce the following abbreviations for compactness
\begin{equation}
  \label{eq:eta-varphi}
  \eta = \sum_{n=1}^{N_p} \frac{M_n}{2 R_n}, \quad
  \quad \varphi = \sum_{n=1}^{N_p} \frac{Q_n}{2R_n}, \quad
  \quad \kappa = 1 + u + \eta \,.
\end{equation}
Therefore, the conformal factor becomes
\begin{equation}
  \label{eq:psi-eta-varphi}
  \psi = \sqrt{\kappa^2 - \varphi^2}\,.
\end{equation}
Equation~\eqref{eq:psi-singular} is essentially an ansatz that states that our
solution is a superposition of Reissner-Nordstr\"om black holes plus corrections
(in $u$), which parallels what is performed in the uncharged case
\cite{Brandt:1997tf}.

Expanding Equation~\eqref{eq:hamiltonian-CTT}, we reach
\begin{multline}
  \label{eq:hamiltonian-CTT-u}
  \kappa \nabla^2 u + \partial_a \kappa \partial^a \kappa - \partial_a \varphi \partial^a \varphi - \partial_a \psi \partial^a \psi
  + \frac{1}{8} \psi^{-6} \bar{A}_{ij}\bar{A}^{ij}
  \\
 + 2 \pi \psi^{-2} \bar{\mathcal{E}} = 0\,.
\end{multline}
In deriving the last expression, we used the fact that the Laplacian
of $\eta$ and $\varphi$ is zero. Equation~\eqref{eq:hamiltonian-CTT-u}
is a second order, non-linear elliptic partial differential equation
in $u$ that depends on $V_{\text{EM}}^i$ through the term
$\bar{A}_{ij} \bar{A}^{ij}$. Now, the momentum and Hamiltonian
constraints~\eqref{eq:constraints-ADM} have been re-expressed as
elliptic
equations~\eqref{eq:momentum-CTT-2},~\eqref{eq:hamiltonian-CTT-u} for
$V_{\text{EM}}^i$ and $u$. The associated boundary conditions are
found from the assumption of asymptotic flatness so that $u$ and
$V_{\text{EM}}^i$ have to go to zero at spatial infinity. In this
paper we assume that $u$ and $V_{\text{EM}}$ are regular everywhere,
and thus they can be found with standard numerical methods that can
solve Equations~\eqref{eq:momentum-CTT-2},~\eqref{eq:hamiltonian-CTT-u}.

The problem of generating valid initial data for multiple charged
black holes is now reduced to solving
Equations~\eqref{eq:momentum-CTT-2} and~\eqref{eq:hamiltonian-CTT-u},
which is done once Maxwell-compliant electromagnetic fields are
found. In this paper, we assume that each puncture is endowed with a
Reissner-Nordstr{\"o}m electromagnetic field, and hence the total
conformal electric field is a superposition of Reissner-Nordstr{\"o}m
electromagnetic fields in isotropic coordinates, i.e.,
\begin{equation}
  \label{eq:RNsuper}
  \bar E^i = \sum_{n=1}^{N_p} \frac{Q_n}{R_n^2}\hat R_n^i,
\end{equation}
nwhere $\hat R_n$ is the radial unit vector centered on the $n-$th puncture. In
the case of a single, non-rotating black hole with zero linear momentum, our
choice of Reissner-Nordstr{\"o}m fields exactly produces a spatial slice of that
solution, since the constraints are solved by $V_{\text{GR}}=V_{\text{EM}} = 0$,
and $u = 0$ [so $\psi = \Psi$, where $\Psi$ is given in Equation~\eqref{eq:RN-Psi}]. For
systems of spinning black holes with linear momenta, the superposition of
Reissner-Nordstr{\"o}m fields is a first approximation to the equilibrium
electromagnetic field generated by these configurations. As for the
gravitational fields generated in the puncture approach (and the gauge fields),
we expect that the time evolution will relax our electromagnetic-field initial
data to their quasi-equilibrium values on a light-crossing timescale. An
advantage of choosing Reissner-Nordstr{\"o}m electromagnetic fields is that they
allow for a clear description of each black hole in the system with a specific
charge, whose isolated horizon value $Q_{\mathcal{S}}$ equals the ``bare''
charge entering Equation~\eqref{eq:RNsuper}. In addition, since there is no
magnetic field, the source term of Equation~\eqref{eq:momentum-CTT-2} vanishes,
and so $V_{\text{EM}} = 0$ (even for multiple black holes with linear and
angular momenta). Thereby, this choice ensures that there are no electromagnetic
contributions to the extrinsic curvature, implying that the parameters entering
Equation~\eqref{eq:V-GR} can be still interpreted as $P_{\text{ADM}}^i = P^i$
and $J_{\text{ADM}}^i = S^i$.

The choice of Reissner-Nordstr{\"o}m electromagnetic fields is by no means unique.
Another possibility is Kerr-Newman fields in quasi-isotropic coordinates. We
present a detailed discussion of this case and the complexities associated with
it in Appendix~\ref{sec:kerr-newm-spac}.

\section{Numerical implementation}
\label{sec:numer-impl}

We implement the formalism outlined in the previous Sections by modifying the
\texttt{TwoPunctures} \cite{Ansorg2004} and \texttt{QuasiLocalMeasures}
open-source codes \cite{Dreyer2003}. The software is run within the
\texttt{Cactus} infrastructure \cite{Goodale:2002a} and all physical
variables are interpolated on a \texttt{Carpet} grid \cite{Schnetter:2003rb,
  Paschalidis2013b}. Black hole apparent horizons are found with
\texttt{AHFinderDirect} \cite{Thornburg2004}.

The main component in our software stack is \texttt{TwoChargedPunctures}, which
is used to generate initial data for two punctures located at $(\pm b, 0, 0)$
given the bare black hole properties ($M_n$, $Q_n$, $P^i_n$, $S^i_n$).
This code implements a pseudo-spectral collocation method that solves the
constraint equations~\eqref{eq:momentum-CTT-2} and \eqref{eq:hamiltonian-CTT-u}
to find $u$ and $V_{\text{EM}}^i$.

In what follows, we adopt Reissner-Nordstr{\"o}m electromagnetic
fields. Since there is only an electric field, $\bar S^i = 0$ in
Equation~\eqref{eq:momentum-CTT-2}, and the momentum constraint is
trivially satisfied by $V_{\text{EM}}^i = 0$.  Hence, we only need to
solve the Hamiltonian constraint~\eqref{eq:hamiltonian-CTT-u}.

\texttt{TwoChargedPunctures} implements a single domain
pseudo-spectral method that covers all $\mathbb{R}^3$ with spatial
infinity on the grid. This region is parametrized by the coordinates
$(A,B,\phi)$, with $A,B \in [-1,1]$ and $\phi \in [0, 2 \pi]$. To be
more specific, the code uses a system of bispherical coordinates that
transform to the usual Cartesian ones with the law\footnote{This parametrization is
  slightly different compared to what is done in
  \cite{Ansorg2004}. The spectral expansion used here treats $A$ and
  $B$ on equal footing, i.e., the spectral decomposition in $A$ and
  $B$ uses the same Chebyshev polynomial basis, unlike what is reported in \cite{Ansorg2004}.}
\begin{subequations}
    \label{eq:twopuncture-coordinates}
  \begin{align}
    x &= b \frac{(1 + A)^2 + 4}{(1 + A)^2 - 4} \frac{2B}{1 + B^2}\,, \\
    y &= b \frac{4 \left(1 + A\right)}{4 - (1 + A)^2}  \frac{1 - B^2}{1 + B^2}
        \cos \phi\,, \\
    z &= b \frac{4 \left(1 + A\right)}{4 - (1 + A)^2}  \frac{1 - B^2}{1 + B^2}
        \sin \phi\,,
  \end{align}
\end{subequations}
where the $x$ axis is along the line connecting the two punctures.
Equations~\eqref{eq:twopuncture-coordinates} describe a set of cylindrical-like
coordinates around the $x$ axis with a radius that depends on both $A$ and $B$.

The coordinates $(A,B,\phi)$ live on a compact grid where spatial infinity
corresponds to $A = 1$, which makes it straightforward to impose the desired
outer boundary conditions ($u \to 0$ at infinity). This condition is enforced by
solving the equations for an auxiliary variable $\mathpzc{U}$ defined as
$u = (A - 1)\mathpzc{U}$. The code expands $\mathpzc{U}$ in Chebyshev
polynomials along $A$ and $B$, and adopts a Fourier basis along $\phi$. The
coordinates are discretized with $n_A$, $n_B$ and $n_\phi$ grid points chosen as
the zeros of Chebyshev polynomials $T_{n_A}(x)$, $T_{n_B}(x)$ and of the sine
function $\sin{(n_\phi \phi)}$. The coefficients of the spectral expansion are found
by evaluating the relevant equation on the collocations points and solving the
corresponding multidimensional non-linear system with a modified Newton-Raphson
method \cite{Barrett1994} (more details on how this is done can be found in
Section II of the original paper \cite{Ansorg2004}). We consider the equations to
be solved, when the residuals are smaller than a threshold
value. To choose this threshold value, we solve for increasingly smaller values
of this threshold and compute the ADM and the horizon masses. When these masses
have converged to within one part in $10^6$, we consider the solution
converged.

With the equations solved and $u$ known, \texttt{TwoChargedPunctures}
reverts back to the physical fields using
Equations~\eqref{eq:K-traceless-transverse},
\eqref{eq:conformally-flat-em-quantities} \eqref{eq:V-decomposition},
\eqref{eq:V-GR} and \eqref{eq:psi-singular}. We then spectrally
interpolate the physical fields on a \texttt{Carpet} grid where
\texttt{AHFinderDirect} is subsequently run to locate the apparent
horizons. Once the horizons are found, we compute mass, charge and
angular momentum of each black hole with our version of
\texttt{QuasiLocalMeasures}, which we call
\texttt{QuasiLocalMeasuresEM}, and which implements the formalism of
isolated horizons for the full Einstein-Maxwell theory as reviewed
in Section~\ref{sec:dynamical-horizons}.  Moreover, having the
spectral expansion of the fields we can interpolate them at a very
large radius to compute the ADM mass, the linear and angular momenta.

\subsection{Code validation}
\label{sec:code-validation}

We validate our approach and numerical implementation with a series of
tests that are presented in this section.

We report our results in terms of the input bare mass $M$ of the punctures,
which is the only mass known a priori.
In all the runs, we confine the black hole in
a region where the \texttt{Carpet} grid resolution is
$\Delta x_i = \num{0.0078}~M$, which usually guarantees that the diameter of the
horizon is resolved by about $100$ points, making it easily found by
\texttt{AHFinderDirect}. We also fix the resolution of the
\texttt{AHFinderDirect} grid to be $79$ points in the azimuthal direction and
$39$ in the meridional direction. We have confirmed that the resolution on the
\texttt{AHFinderDirect} grid has negligible impact in our results. In the cases
presented here, doubling the \texttt{AHFinderDirect} grid resolution introduces
a variation in the computed parameters of order $\SI{0.01}{\percent}$. We compute ADM
integrals by spectrally interpolating our fields on a sphere of radius
$10000~M$, and discretized with $256$ points in both the meridional and
azimuthal directions.

As a first test, we made sure that our modified code with zero charge,
\texttt{TwoChargedPunctures}, produces the same output as the standard
open-source \texttt{TwoPunctures} code. This is not a trivial test because
the equations used in our code and in the original one are different,
having different numerical properties, even though they are
mathematically equivalent. In particular, our formulation is more
susceptible to numerical instabilities due to the finite-arithmetic
error in regions close to the puncture. The reason for this is that
our equations have terms that are not present in the original code,
but that should perfectly cancel out when $Q = 0$. Such a numerical
cancelation near the punctures is not trivial. However, the result of
the test with different spectral resolutions shows that the two
implementations agree at the round-off-error level for punctures with
no charge.

Another key test that our code successfully passes consists in recovering the
only conformally flat analytical solutions known: the Reissner-Nordstr{\"o}m and the
case of two black holes with the same charge-to-mass ratio (see
Appendix~\ref{sec:majumdar-papapetrou} for more details), both of which are
found with $u = 0$. We find the solution $u = 0$ is recovered to machine
precision everywhere outside the horizons, and it is non-identically zero only
very close to the punctures, again due to numerical precision.

\begin{figure}[htbp]
  \centering
  \includegraphics{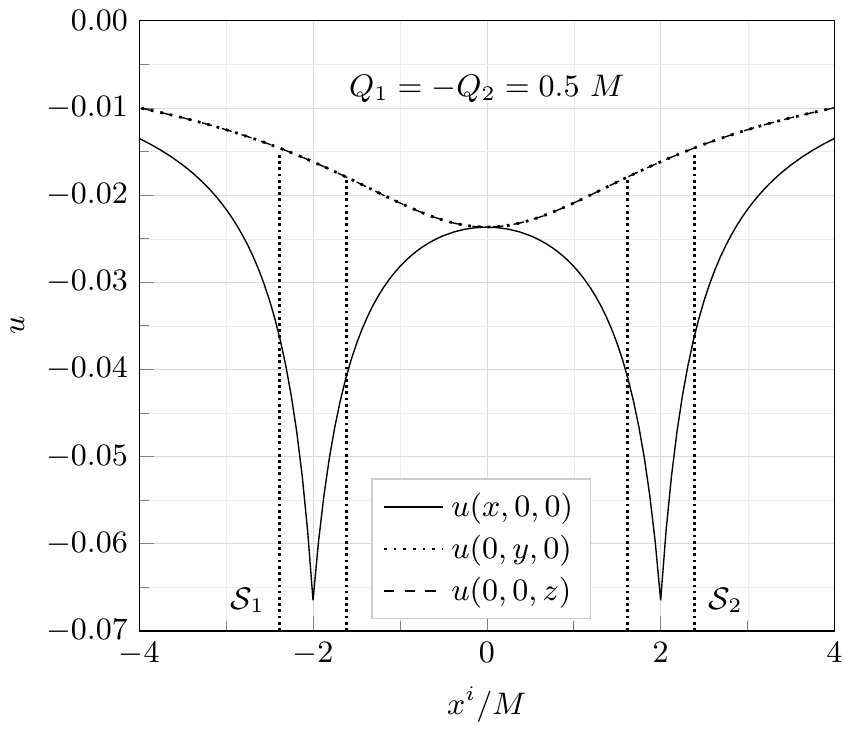}
  \caption{$u$ along the different coordinate axes (solid line for the
    $x$ axis, dotted and dashed for the $y$ and $z$, respectively) for
    two punctures with equal mass and opposite charge $Q_1 = - Q_2 =
    \num{0.5}~M $ located on the $x$ axis at $\pm 2~M$. This
    configuration is generated with a spectral grid resolution $n_A =
    n_B = n_\varphi = 64$. We graphically compared our solution to
    the one in \cite{Alcubierre2009}, and found that the curves
    shown here perfectly match the solution of
    \cite{Alcubierre2009}. The horizons have areal radius
    $R_{\mathcal{S}_1} = R_{\mathcal{S}_2} = \num{0.387}~M$ as defined
    by Equation~\eqref{eq:isolated-radius}. The vertical dotted lines
    represent the coordinate radius of the horizons as found by
    \texttt{AHFinderDirect}. We note that the black hole horizons in
    binary black holes are generally non-spherical, see,
    e.g.,~\cite{PhysRevD.74.064016}}
  \label{fig:puncture-u}
\end{figure}

The next test for \texttt{TwoChargedPunctures} is reproducing the numerical
solution found by \cite{Alcubierre2009} for two non-rotating black holes with
opposite charge-to-mass ratio starting at rest. Figure~\ref{fig:puncture-u}
reports the value of $u$ along the $x$, $y$ and $z$ axes for a system of two
punctures with the same mass but opposite charge ($Q_1 = - Q_2 = \num{0.5}~M$).
We graphically superposed our plot with Figure~1 in \cite{Alcubierre2009},
finding perfect agreement.

\begin{figure}[htbp]
  \centering
  \includegraphics{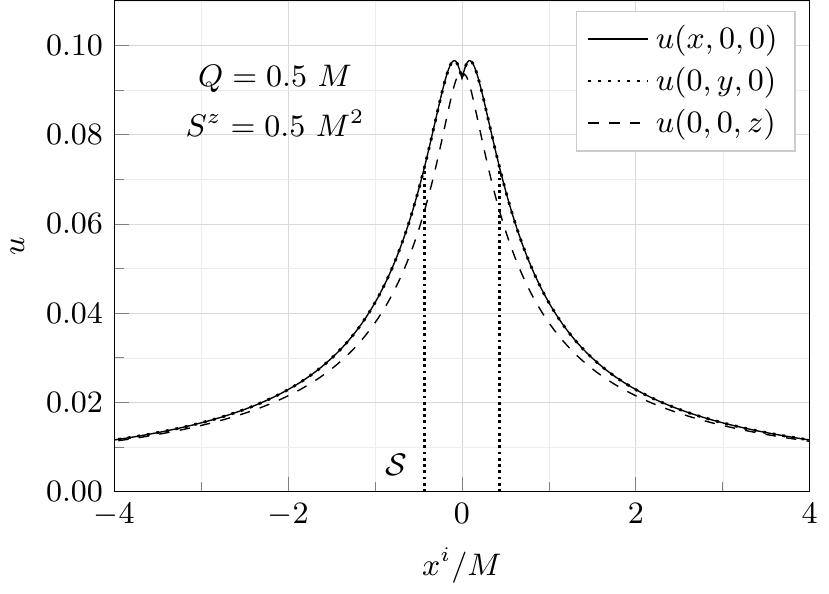}
  \caption{$u$ along the coordinate axes for a single puncture with
    charge $Q = \num{0.5}~M$ rotating around the $z$ axis with angular
    momentum $S^z = \num{0.5}~M^2$. The plot corresponds to spectral grid resolution $n_A
    = n_B = n_\phi = 64$. The horizon has areal radius $R_{\mathcal{S}} =
    \num{0.433}~M$. The different styles of curves have the same meanings
    as in Figure~\ref{fig:puncture-u}.}
  \label{fig:new-puncture-u-spin}
\end{figure}

\begin{figure}[htbp]
  \centering
  \includegraphics{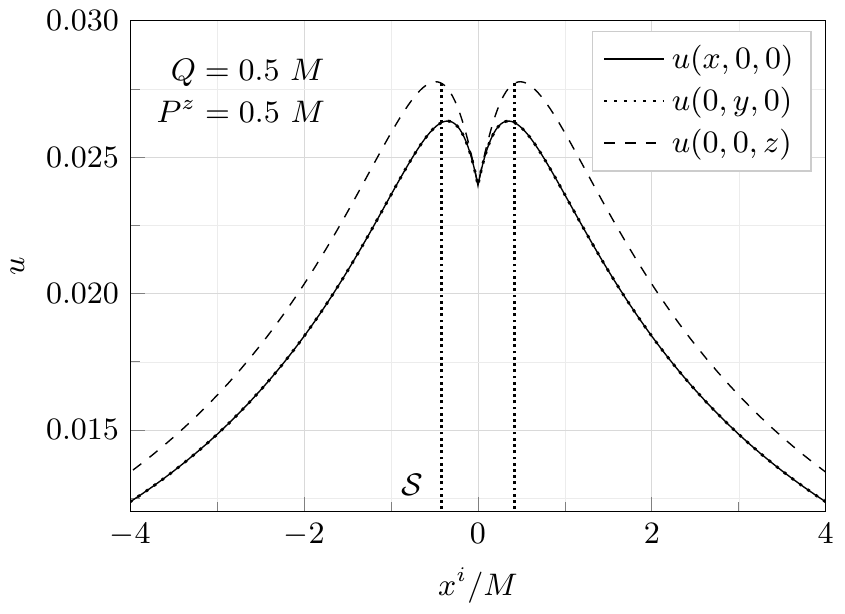}
  \caption{$u$ along the coordinate axes for a single puncture with charge
    $Q = \num{0.5}~M$ with linear momentum
    $P^z = \num{0.5}~M$. The plot corresponds to spectral grid resolution
    $n_A = n_B = n_\phi = 64$. The horizon has areal radius
    $R_{\mathcal{S}} = \num{0.421}~M$. The different styles curves have the same meanings
    as in Figure~\ref{fig:puncture-u}.}
  \label{fig:new-puncture-u-momentum}
\end{figure}

Continuing the progression of complexity in the considered systems, we generate
a single puncture with angular momentum, but no linear momentum, and one with
linear momentum but no angular momentum (Figures~\ref{fig:new-puncture-u-spin}
and \ref{fig:new-puncture-u-momentum}, respectively). In these single-black hole
cases, we compare the horizon mass with the ADM mass measured at infinity and we
find agreement of order $\SI{0.1}{\percent}$ even with resolution as low as $n = 16$.
The same is true for the
ADM angular momentum and the horizon spin, as computed with
\texttt{QuasiLocalMeasuresEM}. We repeated these two tests by aligning the
linear and angular momentum vectors once along the $x$ direction and once along the $z$
direction to ensure that the built-in asymmetry in the coordinates
[Equations~\eqref{eq:twopuncture-coordinates}] does not spoil expected
symmetries in symmetric configurations. By doing this, we find that the
solutions are rotationally invariant to better than one part in $10^6$ for a
resolution $n=32$ or higher.


\subsection{Convergence}
\label{sec:convergence}

\begin{figure}[htbp]
  \centering
  \includegraphics[width = \linewidth]{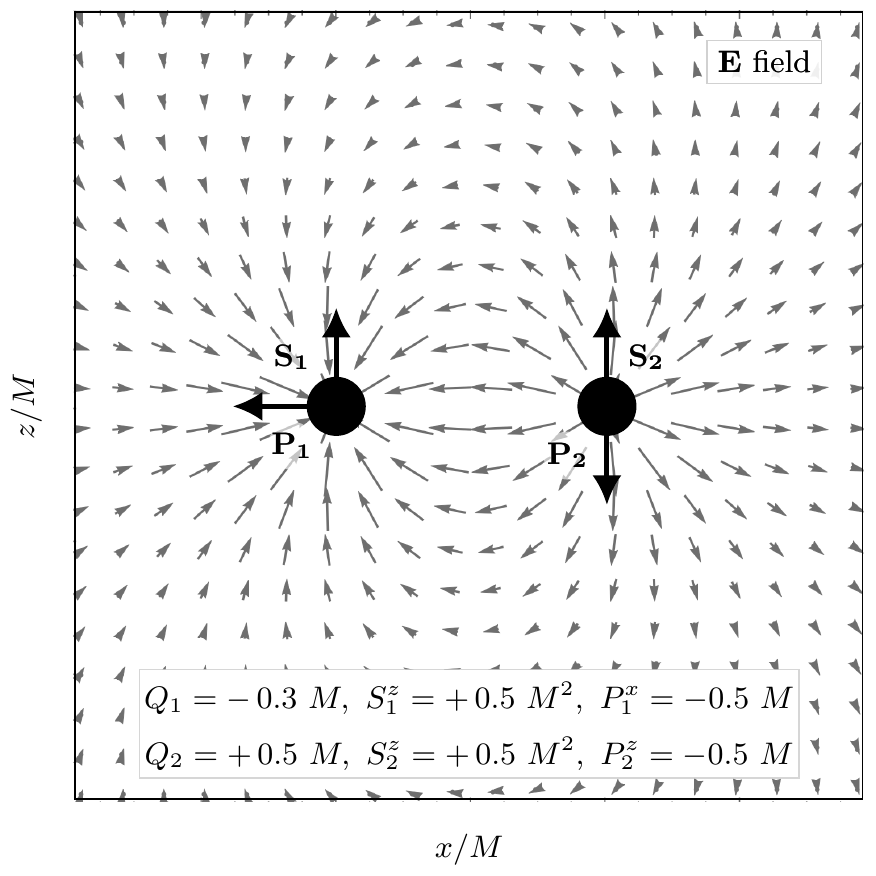}
  \caption{Electric field lines on the $x$-$z$ plane for two charged punctures.
    The first (left) black hole has charge $Q_1=-\num{0.3}~M$, linear momentum
    $P^x_1=-\num{0.5}~M$ and spin angular momentum $S^z_1=\num{0.5}~M^2$. The
    second (right) black hole has $Q_2= \num{0.5}~M$, linear momentum
    $P^z_2=-\num{0.5}~M$ and spin angular momentum $S^z_2=\num{0.5}~M^2$. The
    black disks depict the apparent horizon of each black hole, which set the
    scale in the plot. This is the test case used in the self-convergence test
    reported in Figure~\ref{fig:convergence}. }
  \label{fig:electric-field}
\end{figure}

\begin{figure}[htbp]
  \centering
  \includegraphics{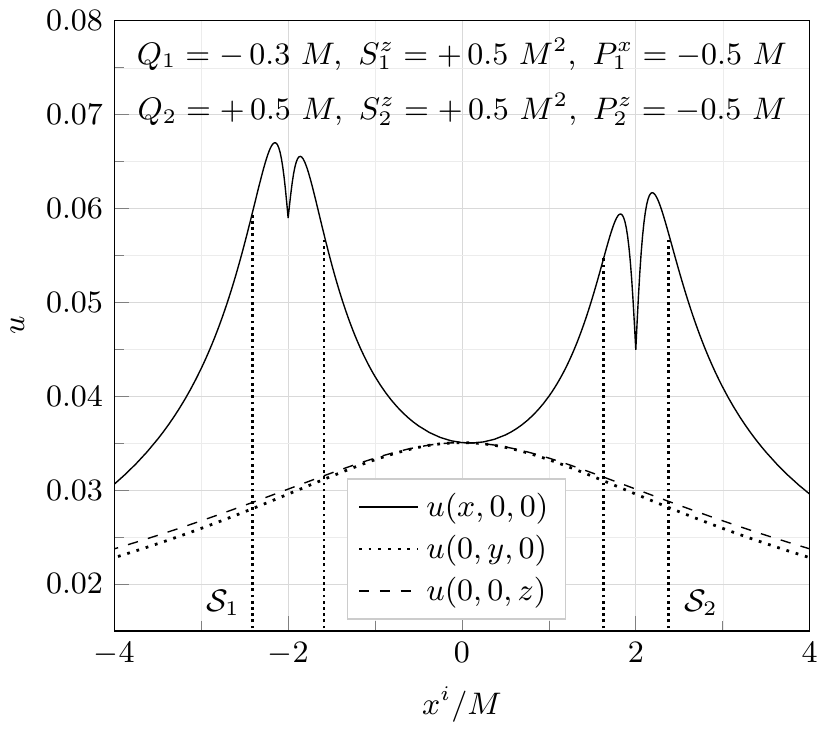}
  \caption{$u$ along the coordinate axes for two punctures with charge
    $Q_1 = -\num{0.3}~M$ and $Q_2 = \num{0.5}~M$. The first black hole has
    linear momentum $P^x_1=-\num{0.5}~M$ and spin angular momentum
    $S^z_1=\num{0.5}~M^2$. The second black hole has linear momentum
    $P^z_2=-\num{0.5}~M$ and spin angular momentum $S^z_2=\num{0.5}~M^2$. The
    electric field lines are reported in Figure~\ref{fig:electric-field}. The
    plot corresponds to spectral grid resolution $n_A = n_B = n_\phi = 64$. This
    system is used for the self-convergence test in
    Figure~\ref{fig:convergence}. The horizons have radii
    $R_{\mathcal{S}_1} = \num{0.412}~M$ and $R_{\mathcal{S}_2} = \num{0.373}$,
    and quasilocal masses $M_{\mathcal{S}_1} = \num{1.187}~M$ and
    $M_{\mathcal{S}_2} = \num{1.202}~M$. The different styles of curves have the
    same meanings as in Figure~\ref{fig:puncture-u}.}
  \label{fig:puncture-u-convergence}
\end{figure}

Finally, we considered the generic system shown in
Figure~\ref{fig:electric-field}. This is formed by two equal-mass
black holes with charge $Q_1 = - \num{0.3}~M$ and $Q_2 =
\num{0.5}~M$. Both black holes are spinning with angular momentum
$S^z_1 =S^z_2= \num{0.5}~M^2$. The black holes also have linear
momentum $P_1^x = P_2^z = -\num{0.5}~M$. The solution for $u$
for this system is depicted in
Figure~\ref{fig:puncture-u-convergence}. With
\texttt{QuasiLocalMeasuresEM}, we find that the quasi-local angular
momenta (charges) agree with their bare counterparts to within one part
in $10^{4}$ ($10^{8}$). We find that the mass of the first horizon
is $\num{1.187}~M$ and the second is $\num{1.202}~M$. The total (ADM)
mass of the system is $\num{2.337}~M$, and the difference between this
value and the sum of the individual masses is the binding energy plus
contribution from ``junk'' radiation.

This system is used to study the self-convergence properties of the code. In
particular, we consider the maximum relative error of $u$ with respect to a
reference solution at high resolution $N$. For this, we sampled $u$ on a set of
points $\mathcal{T}$ and computed the infinity norm
\begin{equation}
  \label{eq:delta-u}
  \|\Delta_n^N u\|_{\infty}^\mathcal{T} = \max_{\vec{x} \in \mathcal{T}}
  \biggr \lvert \frac{ u^n(\vec{x}) - u^N(\vec{x})} {u^N(\vec{x})} \biggr\rvert\,.
\end{equation}
We choose $\mathcal{T}$ as the set of points where spheres of radii $1~M$,
$2~M$, $5~M$, $10~M$, $100~M$, and $1000~M$ intersect the coordinate
axes for $x>0,\, y>0,\, z>0$.

\begin{figure}[htbp]
  \centering
  \includegraphics{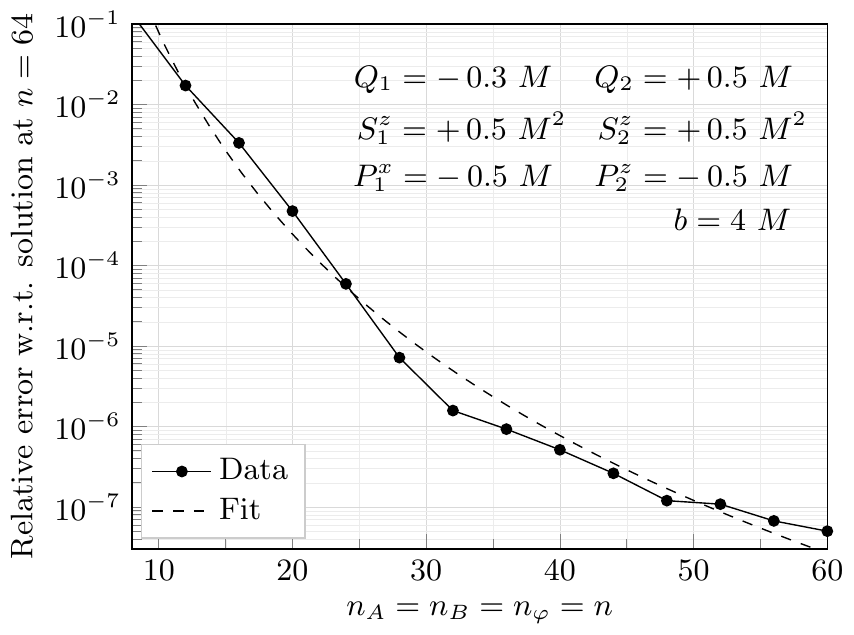}
  \caption{Convergence properties of the algorithm
    measured by computing the maximum relative error on $u$
    $\|\Delta^{64}_{n} u\|_{\infty}^{\mathcal{T}}$ over the test set
    $\mathcal{T}$ (formed by points at distances $1~M$, $2~M$, $5~M$,
    $10~M$, $100~M$, and $1000~M$ on the different coordinate axes). See
    Figure~\ref{fig:electric-field} for the geometric setup and black
    hole parameters used for this test. The dashed line shows that the
    code is approximately sixth-order convergent. All the other
    physical properties (such as the ADM mass, and momenta and the
    horizon quantities) inherit this excellent convergence behavior
    from $u$.}
  \label{fig:convergence}
\end{figure}

We set as a reference solution ($N$) one obtained at high-resolution with
$n_A=n_B=n_\phi=n = 64$, which is between $n=50$ and $n=70$ that were used for
self-convergence tests in the original {\tt TwoPunctures}
code~\cite{Ansorg2004}. Here, we simply choose resolutions which are multiples
of 4, but our results do not depend on this choice. Our convergence test
(Figure~\ref{fig:convergence}) shows that the algorithm is robust; $u$ quickly
converges to its high-resolution value.
The code converges approximately at sixth-order. We also verified that
the code exhibits the same convergence properties when we repeat the
convergence test with $Q_1 = Q_2 = 0$, which also agree with the
convergence properties of the original \texttt{TwoPunctures}
code~\cite{Ansorg2004}. The convergence of $u$ also results in
excellent convergent behavior for both the ADM mass and momenta and
the horizon properties as computed by \texttt{QuasiLocalMeasuresEM}.

\section{Conclusions and future work}
\label{sec:concl-future-work}

Gravitational waves offer new opportunities to study the Universe that
are not accessible with electromagnetic or neutrino astronomy. In this
landscape, numerical-relativity simulations are a powerful tool to
gain insight into the properties and the characteristics of both the
waves and their sources. The majority of numerical-relativity
simulations of black holes to-date do not treat the electric
charge. This is because it is believed that astrophysically relevant
black holes should have a charge which is negligibly small compared to
the mass. For this reason, there are no studies of highly dynamical
electrovacuum spacetimes that involve the inspiral and merger of
binary black holes with charge and spin. Nevertheless, electrovacuum
spacetimes are of great interest, having both a theoretical appeal and
exotic astrophysical applications.

In this paper, we initiated an effort toward solving the coupled
Einstein-Maxwell equations in a dynamical and fully general relativistic regime.
The first step to perform this type of simulations is the generation of valid
initial data. Here, we employed the conformal transverse-traceless approach to
build a formalism for generating initial data for multiple black holes with
charge, angular and linear momenta. Moreover, we applied the theory of isolated
horizons to attribute the physical mass, charge and angular momentum to the
horizon, providing a solid understanding of the physical content of our initial
data. We implemented the formalism in a software based on the
\texttt{TwoPuncture} and the \texttt{QuasiLocalMeasures} open-source codes,
verifying our implementation with a series of tests involving analytical or
previously-known results. The algorithm was found to recover the expected
solutions and showed excellent convergence properties.

With the valid initial data for charged, rotating and moving punctures
it is now possible to simulate dynamical evolution of several systems
that have never been taken in consideration, such as
ultra-relativistic head-on collision, and the quasi-circular or
eccentric inspiral and merger of two black holes. As a first application
of the formalism outlined in this paper we plan to study in the
near-future the case of charged and spinning black holes in
quasi-circular orbit. Some of these simulations are already underway,
and will be presented in forthcoming work.

\begin{acknowledgments}
  We are indebted to the authors of the open-source software that we
  used: \texttt{Cactus}, \texttt{Carpet}, \texttt{TwoPunctures},
  \texttt{AHFinderDirect} and \texttt{QuasiLocalMeasures}. We thank
  V.\ Cardoso, S.\ Gralla, L.\ Lehner, U.\ Sperhake,
  J.\ R.\ Westernacher-Schneider, and M.\ Zilh{\~a}o for useful
  discussions. Computations were performed on the Ocelote cluster at
  The University of Arizona.
\end{acknowledgments}

\appendix

\section{Algorithm and important equations}
\label{sec:algor-import-equat}

In this Appendix, we sketch the algorithm and summarize the important
equations to generate initial data for $3+1$ evolutions of arbitrary
systems of $\mathcal{N}$ black holes with electric charge, linear and
angular momenta using the conformal transverse-traceless
decomposition. In the following $n$ is used to index the n-th black
hole in the system that is $n \in \{1, \dots, \mathcal{N}\}$. Unless
otherwise specified sums in this Appendix are over all punctures. We
also assume that each black hole is endowed with Reissner-Nordstr{\"o}m
electromagnetic fields $(\bar E^i, \bar B^i)$ associated with electric
charge $Q_n$. The steps in generating the initial data are as follows:

\begin{enumerate}[leftmargin=*,noitemsep]
\item Choose the bare parameters $M_n, Q_n, S_n^i, P_n^i,\vec{x}_n$ for each
  black hole, respectively representing mass, charge, angular momentum, linear
  momentum, and position.
\item Compute the conformal electromagnetic fields $(\bar{E}^j_n, \bar{B}^j_n)$ for each
  black hole. Under the assumption of Reissner-Nordstr{\"o}m fields, we obtain
  \begin{align}
    \label{eq:em-field-app}
    \bar E^j_n &=  \frac{Q_n}{R_n^2}\hat R_n^i\,, \\
    \bar B^j_n &= 0\,,
  \end{align}
  with $R_n = \abs{\vec{x} - \vec{x}_n}$ the Euclidean coordinate
  distance from puncture $n$, and $\hat R_n^i$ the corresponding unit vector. Then,
  superpose the conformal electromagnetic fields of all black holes,
    \begin{align}
    \label{eq:em-field-app2}
    \bar E^j &= \sum \bar E^j_n(Q_n, \vec{x}_n)\,, \\
    \bar B^j &= \sum \bar B^j_n(Q_n, \vec{x}_n)\,.
  \end{align}
\item Solve the inhomogeneous momentum constraint for $V_{\text{EM}}^i$
  \begin{equation}
    \label{eq:mom-app}
    (\nabla^2 V_{\text{EM}})^i + \frac{1}{3} \delta^{ij} \partial_j (\partial_k V^k_{\text{EM}})
  - 8 \pi \bar {S}^i = 0\,,
\end{equation}
with
\begin{equation}
  \label{eq:S-app}
 4\pi \bar{S}^i = \bar{\epsilon}^{ijk} \bar{E}_j \bar{B}_k\,,
\end{equation}
and imposing as a boundary condition that $V_{\text{EM}}^i \to 0$ at spatial infinity.

Given our choice for the electromagnetic fields
[Equation~\eqref{eq:em-field-app}], $\bar{S}^i = 0$, so $V_{\text{EM}} = 0$ is a
solution of the momentum constraint~\eqref{eq:mom-app}.
\item
Compute the total auxiliary vector $V^i$,
\begin{equation}
  \label{eq:V-app}
  V^i = V_{\text{GR}}^i + V_{\text{EM}}^i\,,
\end{equation}
with
\begin{equation}
\label{eq:V-GR-app}
 V_{\text{GR}}^i = \sum \left( - \frac{7}{4} \frac{P^i_n}{R_n}
    - \frac{1}{4} \delta_{jk} x_n^jP_n^k \frac{x_n^i}{R_n^3}
    + \frac{\bar{\epsilon}^i_{jk} x_{n}^jS_{n}^k}{R_n^3} \right)\,.
\end{equation}
\item
Solve the Hamiltonian constraint for $u$, imposing $u \to 0$ at spatial infinity
\begin{multline}
\label{eq:ham-app}
  \kappa \nabla^2 u + \partial_a \kappa \partial^a \kappa - \partial_a \varphi \partial^a \varphi - \partial_a \psi \partial^a \psi \\
  + \frac{1}{8} \psi^{-6} \bar{A}_{ij}\bar{A}^{ij}
  + 2 \pi \psi^{-2} \bar{\mathcal{E}} = 0\,.
\end{multline}
with 
  \begin{align}
    \label{eq:sym-app}
    \kappa &= 1 + u + \eta\,, \\
    \eta &= \sum \frac{M_n}{2 R_n}\,, \\
    \varphi &= \sum \frac{Q_n}{2R_n}\,, \\
    \psi &= \sqrt{\kappa^2 - \varphi^2}\,, \\
\bar A_{ij} &=  2 V_{(i,j)} - \frac{2}{3} \delta_{ij} \partial_k V^k\,, \\
4\pi \bar{\mathcal{E}} &= \frac{1}{2} (\bar{E}_i \bar{E}^i + \bar{B}_i \bar{B}^i)\,,
  \end{align}

\item  With $\psi$ now known, compute the physical fields that are necessary for the evolution
  \begin{align}
    \label{eq:phys-app}
    E^i    & = \psi^{-6} \bar{E^i}\,, \\
    B^i    & = \psi^{-6} \bar{B^i}\,, \\
    \gamma_{ij} & = \psi^4 \delta_{ij}\,,    \\
    K_{ij} & = \psi^{-2} \left( 2 V_{(i,j)} - \frac{2}{3} \delta_{ij} \partial_k V^k\right)\,.
\end{align}
\item Find the isolated horizons ${\mathcal{S}}_n$ and compute the associated physical properties
  \begin{align}
    \label{eq:iso-app}
    Q_{\mathcal{S}_n} &= \frac{1}{4\pi} \int_{\mathcal{S}_n} {}^{(4)}{}^{\star}F\,,\\
    R_{\mathcal{S}_n} &= \left( \frac{1}{4\pi} \int_{\mathcal{S}_n} \epsilon \right)^{\frac{1}{2}}\,, \\
    J_{\mathcal{S}_n} &= -\frac{1}{8\pi} \int_{\mathcal{S}_n} (\varphi \cdot \omega)\epsilon + 2(\varphi \cdot
                        {}^{(4)}A) {}^{(4)}{}^{\star}F \,, \\
    M_{\mathcal{S}_n} &= \frac{1}{2R_{\mathcal{S}_n}} \left[(R_{\mathcal{S}_n}^2 + Q_{\mathcal{S}_n}^2)^2 + 4J_{\mathcal{S}_n}^2\right]^{\frac{1}{2}}\,,
  \end{align}
  where ${}^{(4)}{}^{\star}F$ is the dual of the electromagnetic tensor,
  $\epsilon$ is the horizon surface $2$-form, ${}^{(4)}A$ is the electromagnetic
  vector potential, $\varphi$ is the approximate rotational Killing vector on
  $\mathcal{S}_n$, and $\omega$ is defined in the main text [see
  Equation~\eqref{eq:isolated-spin}].
  $Q_{\mathcal{S}_n}, R_{\mathcal{S}_n}, J_{\mathcal{S}_n}$ and
  $M_{\mathcal{S}_n}$ are respectively the charge, radius, angular momentum and
  mass of the n-th horizon.

\end{enumerate}
\section{Isolated horizon in the Reissner-Nordstr{\"o}m solution}
\label{sec:reissner-nordstrom-1}

The goal of this Appendix is to show that the formalism of isolated horizons
produces the expected black hole properties in the case of the
Reissner-Nordstr{\"o}m solution. This can be proven starting from
metric~\eqref{eq:rn-usual-metric}, which we rewrite here for convenience
\begin{multline}
  \label{eq:rn-usual-metric2}
  \d s^2 = - \left(1 - \frac{2m}{r} + \frac{q}{r^2}\right) \d t^2
  + \\
  \left(1 - \frac{2m}{r} + \frac{q}{r^2}\right)^{-1} \d r^2 + r^2 (\d \theta^2 +
  \sin^2\theta \d \phi^2)\,,
\end{multline}
with electromagnetic potential
\begin{equation}
  \label{eq:rn-A2}
  {}^{(4)}A = -\frac{q}{r} \d t\,.
\end{equation}
In this case, a spherical surface with coordinate radius $r_+ = m +
\sqrt{m^2 - q^2}$ is a Killing horizon, which implies that it is an
isolated horizon. This is because every Killing horizon which is
topologically $S^2\times \mathbb{R}$ is an isolated horizon
\cite{Ashtekar2000}. Therefore, the metric $\mathbbm{q}_{ab}$ induced
on the spatial section of the horizon is simply the metric on a
spherical surface [$\d s^2= \mathbbm{q}_{ab}\d x^a \d x^b= r^2
(\d \theta^2+\sin^2\theta \d \phi^2)$], and the value of $R_{\mathcal{S}}$
defined by Equation~\eqref{eq:isolated-radius} coincides with $r_+$
itself, since the radial coordinate in
Equation~\eqref{eq:rn-usual-metric2} is the areal radius.  In this
case, the rotational vector $\varphi$ in~\eqref{eq:isolated-spin} is
taken to be the generator of the azimuthal symmetry on the sphere,
which is also a Killing vector of the entire spacetime. Hence, we find
that $\varphi \cdot {}^{(4)}A = 0$ as ${}^{(4)}A$ has only a temporal
components and $\varphi$ only spatial. Moreover, since the
future-directed vector $k^a$ orthogonal to $\mathcal{S}$ has only
radial and temporal component, and any $t^a$ tangent to $\mathcal{S}$
has only azimuthal and meridional components, $t^a \nabla_a k^b =
0$. By construction, we also have $t^a \nabla_a k^b = t^a \omega_a k^b
= 0$, which implies that $\omega_a = 0$, because the equation is zero
for each $t^a$. Hence, by use of Equation~\eqref{eq:isolated-spin}, we
conclude that $J_{\mathcal{S}} = 0$.

To compute charge and mass, we need the electromagnetic tensor, which is given by
\begin{equation}
  \label{eq:rn-F}
  {}^{(4)}F = \d {}^{(4)}A = -\frac{q}{r^2} \d r \wedge \d t = \frac{q}{r^2} \d t \wedge \d r \,,
\end{equation}
and its dual
\begin{equation}
  \label{eq:rn-F-star}
  {}^{(4)}{}^{\star}F =  \sqrt{-g}  \frac{q}{r^2} \d \theta \wedge \d \varphi = q \sin \theta  \d \theta \wedge \d \varphi  \,.
\end{equation}
The integration of ${}^{(4)}{}^{\star}F \slash 4 \pi$ over any sphere
of coordinate radius $r$ results in exactly $q$, so
Equation~\eqref{eq:isolated-charge} implies $Q_{\mathcal{S}} = q$.

Finally, from Equation~\eqref{eq:isolated-mass} the horizon mass is
\begin{equation}
  \label{eq:mass-h-rn}
  M_S = \frac{(R_S^2 + q^2)}{2R_S} = \frac{2 m (m  +  \sqrt{m^2 - q^2})}{2 (m +
    \sqrt{m^2 - q^2})} = m\,.
\end{equation}
For a Reissner-Nordstr{\"o}m black hole $m$, $q$ are interpreted as the
spacetime total energy and electric charge,
respectively~\cite{Wald1984}. Therefore, in this case, the bare mass
(charge), the isolated horizon mass (charge), and the physical mass
(charge) all coincide.

\section{Computing the charge of an isolated horizon}
\label{sec:charge-isol-horiz}

In this Appendix we discuss how we perform the computation of the
horizon charge.  To compute the charge of the horizon, we need to perform
the following integration (see Section~\ref{sec:dynamical-horizons}):
\begin{equation}
  \label{eq:isolated-charge-2}
  Q_{\mathcal{S}} = \frac{1}{4\pi} \int_{\mathcal{S}} \frac{1}{2!} \epsilon_{abc} E^c \d x^a \wedge \d x^b \,.
\end{equation}
This quantity is coordinate-independent, so choosing Cartesian coordinates
$(x^a) = (x,y,z)$, we can write
\begin{equation}
  \label{eq:star-F-cartesian}
  Q_{\mathcal{S}} = \frac{1}{4\pi} \int_{\mathcal{S}} \sqrt{\gamma} \left(
    E_z \d x \wedge \d y + E_x \d y \wedge \d z - E_y \d x \wedge \d z \right)\,,
\end{equation}
with $\gamma$ determinant of the spatial metric. We introduce a parametrization of
$\mathcal{S}$ with polar coordinates $(\theta,\phi)$ around the origin $(x_0, y_0, z_0)$,
\begin{equation}
  \label{eq:parametrization-s}
  \begin{cases}
    x(\theta, \phi) = x_0 + s(\theta, \phi) \sin\theta \cos\phi \\
    y(\theta, \phi) = y_0 + s(\theta, \phi) \sin\theta \sin\phi \\
    z(\theta, \phi) = z_0 + s(\theta, \phi) \cos\theta
  \end{cases}
  \,,
\end{equation}
with $s(\theta,\phi)$ suitable smooth function. This is always possible since by
hypothesis $\mathcal{S}$ has spherical topology and by construction
$Q_{\mathcal{S}}$ does not depend on the parametrization. Then, the first term
in Equation~\eqref{eq:star-F-cartesian} can be written as
\begin{multline}
  \label{eq:integral-F-xy}
  \int_S  \sqrt{\gamma} E_z(x,y,z) \d x \wedge \d y \\= \int_\theta \int_\phi  \sqrt{\gamma} E_z(\theta, \phi) \lvert \det J_{xy}(\theta,\phi) \rvert \d \theta \d \phi\,,
\end{multline}
where $J_{xy}(\theta, \phi)$ is the Jacobian of the
transformation~\eqref{eq:parametrization-s} involving the coordinates $x$ and
$y$
\[
  J_{xy}(\theta, \phi) =
\begin{pmatrix}
    \partial_\theta x(\theta, \phi)      & \partial_\phi x(\theta, \phi)  \\
    \partial_\theta y(\theta, \phi)      & \partial_\phi y(\theta, \phi)  \\
  \end{pmatrix}
  \,.
\]
The remaining terms in Equation~\eqref{eq:star-F-cartesian} are dealt with
accordingly.

In \texttt{QuasiLocalMeasuresEM}, we use the parametrization
$s(\theta, \phi)$ provided by \texttt{AHFinderDirect}, and we compute
the derivatives in the Jacobians using a centered, second-order
accurate finite-difference scheme.

\section{Kerr-Newman spacetime}
\label{sec:kerr-newm-spac}

In this Appendix we review the Kerr-Newman spacetime and discuss
challenges associated with using the Kerr-Newman electromagnetic
fields as source terms in the Hamiltonian and momentum constraints.

The Kerr-Newman black hole with mass $m$, electric charge $q$, and angular
momentum $a m$ in Boyer-Lindquist coordinates
$(t, r, \theta, \phi)$ is \cite{Wald1984}
\begin{equation}
  \label{eq:Kerr-Newman-ds}
  \begin{split}
    \d s^2 =& -\frac{\Delta - a^2 \sin^2 \theta}{\rho^2} \d t^2  + \frac{\rho^2}{\Delta} \d
    r^2 +  \rho^2 \d \theta^2 \\
    &- 2 a \sin^2 \theta \frac{(r^2 + a^2
      -\Delta)}{\rho^2} \d t \d \phi                         \\
    &+ \frac{(r^2 + a^2)^2 - \Delta a^2 \sin^2 \theta}{\rho^2} \sin^2 \theta \d \phi^2\,,
  \end{split}
\end{equation}
with
\begin{subequations}
  \label{eq:Kerr-Newman-symbols}
  \begin{align}
    \rho^2 & = r^2 + a^2 \cos^2 \theta\,, \\
    \Delta   & = r^2 - 2mr + a^2 + q^2\,.
  \end{align}
\end{subequations}
The electromagnetic vector potential is
\begin{equation}
  \label{eq:Kerr-Newman-em}
  {}^{(4)}A = - \frac{qr}{\rho^2}(\d t - a \sin^2 \theta \d \phi)\,.
\end{equation}
Following the usual procedure for generating puncture initial data, we transform
to quasi-isotropic coordinates by introducing a new radial coordinate $R$ as in \cite{Zilhao2014}
\begin{equation}
  \label{eq:R-quasi-isotropic}
  r = R \left(1 + \frac{m}{R} + \frac{R_H^2}{R^2}\right)\,,
\end{equation}
with
$R_H = \frac{1}{2} \sqrt{m^2 - a^2 - q^2}$
radius of the black hole horizon in the new coordinate system. The metric takes
now the form
\begin{equation}
  \label{eq:Kerr-Newman-quasi-isotropic}
  \d s^2 = (- \alpha^2 + \beta_\phi \beta^\phi) \d t^2 + 2 \beta_\phi \d \phi \d t + \gamma_{lk} \d x^l \d x^k\,,
\end{equation}
where
\begin{equation}
  \label{eq:Kerr-Newman-spatial}
  \begin{split}
    \gamma_{lk} \d x^l \d x^k = \Psi^4 [ &\d R^2 + R^2\d \theta^2 +  R^2 \sin^2 \theta \d\phi^2 \\
    &\;  a^2 h R^4 \sin^4 \theta \d\phi^2 ]\,,
  \end{split}
\end{equation}
where $\Psi$ is the conformal factor, $l, k \in \{R, \theta, \phi\}$, and $\alpha$, $\beta$, $\gamma$ and $h$ functions of $(R,\theta,\phi)$, with
\begin{subequations}
  \begin{align}
    \label{eq:alpha-sigma-quasi-isotropic}
    \Psi^4    & = \rho^2 \slash R^2\,,                                     \\
    \alpha      & = \frac{\rho^6 (R + R_H)(R - R_H)}{R \Upsilon}\,,            \\
    \beta_\phi    & = -a \sigma \sin^2 \theta\,,                                 \\
    \beta^\phi    & = \beta_\phi \slash \gamma_{\phi\phi}\,,                                  \\
    h      & = (1 + \sigma) \slash (\rho^2 R^2)\,,                           \\
    \sigma      & = \frac{2 m r - q^2}{\rho^2}\,,         \\
    \Upsilon      & = \rho^6 \sqrt{r^2 + a^2(1 + \sigma\sin^2\theta)}\,, \\
    \gamma_{\phi\phi} & = \sin^2\theta \left(r^2 + a^2(1 + \sigma \sin^2 \theta)\right)\,.
  \end{align}
\end{subequations}

The non-zero components of the electromagnetic fields are\footnote{
  Our expression for $E^\theta$ differs from the corresponding one in
  Equation~(3.5) of \cite{Zilhao2014} by a factor of $r/R$. We find that the electric field
  components listed in~\cite{Zilhao2014} do not satisfy Maxwell's
  equations, and that Gauss's law yields a value for the charge that
  is correct for spherical surfaces, but the value is different on
  non-spherical surfaces, e.g. ellipsoidal ones. We have checked that
  our electric fields satisfy Maxwell's equations, and, as a result,
  Gauss's law yields the correct electric charge even on non-spherical
  surfaces. We conclude that $E^\theta$ in \cite{Zilhao2014} has a
  typographical error.}
\begin{subequations}
  \label{eq:em-quasi-isotropic}
  \begin{align}
    E^R &= \frac{qR(2r^2 - \rho^2)(r^2 + a^2)}{\Upsilon}\,, \\
    E^\theta &= -\frac{2a^2 q (R - R_H) (R + R_H) r\cos \theta \sin \theta} {R\Upsilon}\,, \\
    B^R &= \frac{2aqRr(r^2 + a^2) \cos \theta}{\Upsilon}\,, \\
    B^\theta &= \frac{a q (R - R_H) (R + R_H) (2 r^2 - \rho^2) \sin \theta} {R \Upsilon}\,.
  \end{align}
\end{subequations}
The conformal fields are obtained by scaling by
$\sqrt{\gamma} = \Upsilon \rho^{-4} R^{-1}\sin\theta$
\begin{subequations}
  \label{eq:em-quasi-isotropic-conformal}
  \begin{align}
    \bar{E}^R &= \frac{q(2r^2 - \rho^2)(r^2 + a^2)\sin \theta}{\rho^4}\,, \\
    \bar{E}^\theta &= -\frac{2a^2 q (R - R_H)(R + R_H) r \cos \theta \sin^2 \theta}{\rho^4 R^2}\,, \\
    \bar{B}^R &= \frac{2aqr(r^2 + a^2) \cos \theta \sin \theta}{\rho^4}\,, \\
    \bar{B}^\theta &= \frac{a q (R - R_H) (R + R_H) (2 r^2 - \rho^2) \sin^2 \theta}{\rho^4 R^2}\,.
  \end{align}
\end{subequations}
In these coordinates, the conformal fields are regular for $R \to 0$ (in this limit
$\rho \sim r \sim 1/R$).

However, Equations~\eqref{eq:momentum-CTT-2} and~\eqref{eq:hamiltonian-CTT-u} are in
Cartesian coordinates. Transforming to Cartesian coordinates as in flat spacetime, the conformal
fields are obtained as
\begin{subequations}
  \label{eq:em-quasi-isotropic-conformal-cartesian}
  \begin{align}
    \bar{E}^i &= \frac{\partial x^i}{\partial R} \frac{\bar{E}^R}{R^2 \sin \theta} + \frac{\partial x^i}{\partial
                \theta} \frac{\bar{E}^\theta}{R^2 \sin \theta} \,, \\
    \bar{B}^i &= \frac{\partial x^i}{\partial R} \frac{\bar{B}^R}{R^2 \sin \theta} + \frac{\partial
                    x^i}{\partial \theta} \frac{\bar{B}^\theta}{R^2 \sin \theta} \,,
  \end{align}
\end{subequations}
where here $i \in \{x, y, z\}$ and the factor of $R^2 \sin \theta$ is the determinant of the
Jacobian of the transformation and ensures that the resulting fields $\bar{E}^i$
and $\bar{B}^i$ satisfy the Maxwell constraints
\begin{subequations}
  \label{eq:em-constraints-flat}
  \begin{align}
    \partial_i \bar{E}^i &= 0\,, \\
    \partial_i \bar{B}^i &= 0 \,.
  \end{align}
\end{subequations}
In these coordinates, the fields are singular when $x,y,z \to 0$. Given this
singular behavior, $V_{\text{EM}}^i$ is expected to be singular as well near the
punctures because the source of the momentum
constraint~\eqref{eq:momentum-CTT-2} diverges with a high power of R. This is
precisely what we find when we implement our algorithm with the Kerr-Newman
electromagnetic fields. In particular, for a single Kerr-Newman black hole
without linear momentum, the singular source terms are $\bar{S}^x$ and
$\bar{S}^y$, which at leading order for $x,y,z\to0$ scale as
\begin{subequations}
  \label{eq:Sphi-R0}
  \begin{align}
    \bar S^x &\sim \frac{aq^2 y}{R_H (x^2 + y^2 + z^2)^{\frac{5}{2}}}  \,, \\
    \bar S^y &\sim -\frac{aq^2 x}{R_H (x^2 + y^2 + z^2)^{\frac{5}{2}}}  \,.
  \end{align}
\end{subequations}

A possible approach to dealing with the singular source would be to
separate the singular part of the solution from the regular one, as is
done for the Hamiltonian constraint. However, this approach typically
requires a known analytic solution, and this does not seem possible
within the conformal flatness approximation, because the Kerr-Newman
solution does not admit conformally flat spatial slices. In future
work, we will explore potential solutions to these challenges by
lifting the conformal flatness approximation.

\section{Generalized Majumdar-Papapetrou}
\label{sec:majumdar-papapetrou}

In this Appendix we show that our formalism recovers spatial slices of
a generalized Majumdar-Papapetrou's solution found by
\cite{Alcubierre2009} when each black hole is at rest, non-spinning
and all black holes have the same charge-to-mass ratio. This happens
because under these assumptions the momentum constraint is trivially
satisfied, and the Hamiltonian one is solved by $u = 0$, as we verify
in what follows.

Given our definitions of $\eta$ and $\varphi$
[Equations~\eqref{eq:eta-varphi}], if the charge-to-mass ratio is
fixed to $\lambda$ for every black hole, then $\varphi = \lambda
\eta$. Moreover, with our choice of Reissner-Nordstr{\"o}m fields, there
are no magnetic fields, so the electromagnetic energy is $8 \pi
\bar{\mathcal{E}} = 4 \partial_a \varphi \partial^a \varphi$, where
the factor of $4$ arises from the fact that $\varphi$ is not the
electrostatic potential but it is \emph{half} of it. Plugging the
ansatz $u = 0$ into the Hamiltonian constraint
[Equation~\eqref{eq:hamiltonian-CTT-u}] yields
\begin{equation}
  \label{eq:hamiltonian-CTT-equal-ratio}
  \partial_a \kappa \partial^a \kappa - \partial_a \varphi \partial^a \varphi - \partial_a \psi \partial^a \psi + \psi^{-2} \partial_a \varphi \partial^a \varphi = 0\,.
\end{equation}
But, $\psi = \sqrt{\kappa^2 - \phi^2}$, thus, multiplying the last equation by
$\psi^2$, and expressing the derivatives of $\psi$ in terms of $\kappa$, $\phi$ and their
derivatives yields
\begin{multline}
  \label{eq:hamiltonian-CTT-equal-ratio-step2}
  (1 - \kappa^2) \partial_a \varphi \partial^a \varphi - \varphi^2 \partial_a \kappa \partial^a \kappa + 2 \kappa \varphi \partial_a \kappa \partial^a \varphi = 0\,.
\end{multline}
Plugging $\kappa =1+ \eta=1 + \varphi \slash \lambda$, and $\partial_a
\kappa = \partial_a\varphi \slash \lambda$ into this last expression,
after some algebra we find that the Hamiltonian constraint is
satisfied. If we choose $\lambda = 1$ we find
\begin{equation}
  \label{eq:majumdar-papapetrou-spatial}
  \gamma_{ij} = \left(1 + \sum_{n=1}^{N_p} \frac{M_n}{R_n} \right) \delta_{ij}\,,
\end{equation}
which describes a spatial slice of the Majumdar-Papapetrou spacetime with $N$
extremal black holes \cite{Majumdar1947, Papapetrou1945}.

\bibliographystyle{apsrev4-1}
\bibliography{bibliography,einsteintoolkit}

\end{document}